\documentclass[10pt,a4paper,preprint,single-columns]{revtex4-1}

\usepackage{graphicx}
\usepackage[breaklinks=true,colorlinks=true,linkcolor=blue,urlcolor=blue,citecolor=blue]{hyperref}
\usepackage{soul}
\setstcolor{red}

\usepackage{amsfonts,amssymb,amsthm,array}

\usepackage{verbatim}

\expandafter\let\csname equation*\endcsname\relax

\expandafter\let\csname endequation*\endcsname\relax

\usepackage{amsmath}
\usepackage{url}

\usepackage{color}

\def\yellow{\color{yellow}}

\theoremstyle{plain}

\theoremstyle{definition}

\newcommand{\al}{\alpha}
\newcommand{\g}{\gamma}
\newcommand{\bz}{\bar z}

\newcommand{\ii}{{\rm{i}}}
\newcommand{\dd}{{\rm{d}}}

\newcommand{\nn}{\nonumber}

\def\sign{\mbox{sign}}
\newcommand{\eq}[1]{(\ref{#1})}
\renewcommand{\>}{\rangle}
\newcommand{\la}{\label}
\newcommand{\ba}{\begin{align}}
\newcommand{\ee}{\end{equation}}
\newcommand{\be}{\begin{equation}}
\newcommand{\IM}{\mbox{Im}\,}
\newcommand{\RE}{\mbox{Re}\,}
\def\12{\frac{1}{2}}
\newcommand{\p}{\partial}
\newcommand{\en}{\end{align}}
\newcommand{\<}{\langle}

\usepackage{color}

\def\yellow{\color{yellow}}

\begin{document}

\title[Geometric Singularities]{Quantum Hall States  and Conformal  Field
 Theory  on a Singular Surface}

\author{T. Can} 
\affiliation{Initiative for the Theoretical Sciences, The Graduate Center, CUNY,
New York, NY 10012, USA}

\author{P. Wiegmann}
\altaffiliation[]{also at IITP RAS, Moscow 127994 Russia}
\affiliation{Kadanoff Center for Theoretical Physics, University of Chicago,
5640 South Ellis Ave, Chicago, IL 60637, USA}




\begin{abstract}

In \cite{Can2016}, quantum Hall states on singular surfaces were shown to
possess an emergent conformal symmetry. In this paper, we develop this idea
further and flesh out  details on the emergent conformal symmetry in holomorphic
adiabatic states, which we define in the paper.  We highlight the connection
between the  universal features of  geometric transport of
quantum Hall  states  and holomorphic dimension of primary fields in  conformal
field theory.  In parallel we  compute the  universal finite-size corrections
to the free energy of a critical system on a hyperbolic sphere with conical
and cusp singularities, thus extending the result of Cardy and Peschel for
critical systems on a flat cone \cite{Cardy1988}, and the known results for 
critical systems  on polyhedra and flat branched Riemann
surfaces.

\end{abstract}

\maketitle
%
%

\tableofcontents

%

\section{ Introduction}\la{Sec1}  This paper is devoted to three subjects:
(i) Geometric transport of quantum Hall states (QH) on singular  surfaces.
These are surfaces  with multiple  conical and parabolic singularities;
(ii) Finite-size corrections to the free energy of critical systems on such
surfaces; (iii) connection between quantum Hall effect (QHE) and critical
systems in two dimensions.

\subsection{ Finite-size corrections in critical systems on singular surfaces}

In the seminal paper \cite{Cardy1988}, Cardy and Peschel   computed  the
finite-size correction  to the free energy of critical systems
on a flat conical surface. The free energy of a critical system   consists of the extensive part which grows with system size,  and a
finite-size correction, the intrinsic part,   which grows at most logarithmically
with the system size. 

  The   extensive part    typically  depends on details of the system at
the smallest (say, lattice) scale, and is  known to be non-universal. However,
the finite-size correction  is geometrical in nature, independent of microscopic
details, and hence  a universal characteristic of the system. It   also represents
the Casimir effect of  one dimensional conformal invariant quantum  field
theory, as well as its specific heat. 

In \cite{Cardy1988}, the free energy was argued to scale logarithmically
in volume  \be (V\p_V)f=-\frac c {12} \chi,\la{109}\ee with a scale-free coefficient
given by the Euler characteristic \(\chi\), a topological invariant of the
surface, and \(c
\) the central charge of the critical system. 
  
 Cardy and
Peschel
 \cite{Cardy1988} also computed the finite-size correction of an isolated
conical, or elliptic, singularity. We denote the deficit angle by  \(2\pi\alpha\),
and the total opening angle, or simply cone angle, by \(2\pi\gamma\), with
$\gamma = 1 - \alpha$ (see Fig.\ref{Fig1}). Then, the intrinsic part of the
 free energy  at \(L\to \infty\) was found to be \begin{align}
(V\p_V)f=-\frac c{24}\frac{h_\g}{\gamma}, \quad h_{\gamma} = 1 - \gamma^{2}
= \alpha(2 - \alpha).
  \la{cardy-cal}  
\end{align}

This formula is valid for all values of $\gamma >0$. When $0< \gamma \le
1$, the conical singularity has positive net curvature concentrated near
the tip, and it can be embedded in 3D as a ``party hat" (see Fig.\ref{Fig1}).
For $\gamma > 1$, the curvature is negative and the opening angle exceeds
$2\pi$. This is the case of a branched Riemann surface. A branched Riemann
surface can be described by a metric with integer $\gamma = n$ singularities
at branch points, since the opening angle indicates that only after $2\pi
n$ traversals around the origin does one return to the starting point. In
this case, \(\alpha=1-n\) and the sign of the free energy correction  changes.

 Conformal field theory on {\it flat} branched Riemann surfaces had been
studied in early works by  Knizhnik \cite{Knizhnik1987} and  related work
 \cite{dixon1985,bershadsky1987,sonoda1987,zamolodchikov1987}.

In \cite{calabrese2004}, Cardy and Calabrese used an extension of the formula
\eq{cardy-cal} for \(\alpha=1-n\) to compute the entanglement entropy in
one-dimensional quantum conformal invariant systems, by identifying the Renyi
entropy with the free energy correction of a system on the \(n\)-sheeted,
also flat, Riemann
surface. The entanglement entropy follows from the limit $n =1$. 

In these works, the singularities or branch points have an interpretation
as primary operators of a conformal field theory, with a holomorphic conformal
dimension given by \(h_\g\). More precisely, this observation implies that
a critical system on a single cone of the linear size $L$ is equivalent to a system on a disk  of  the
size \(L^{\frac 1{\gamma}}\), with a special vertex operator of conformal
dimension $h_{\g}$ inserted in the center of the disk, Fig.\ref{Fig1}.

The  origin of the finite-size energy, or Casimir effect, 
 is the gravitational, or trace, anomaly: despite an apparent scale invariance,
the trace of the stress tensor does not vanish.

\bigskip

Critical systems  on  surfaces with multiple singularities are less studied.
Comprehensive  results are available for  polyhedra, flat surfaces whose
vertices
represent conical singularities. In this case, the essential part of the
spectral determinant has been found explicitly in Refs. \cite{Aurell1,Aurell2,kokotov2011spectral,kokotov2013polyhedral}.

In the present work, we will elaborate further on the properties of  critical
systems on constant curvature (not necessarily flat) surfaces with
many singularities, elliptic and parabolic alike. 

 Extension of the results for critical systems on flat surfaces to singular
positively and negatively curved surfaces meets difficulties related to a
lack of explicit formulas in the uniformization theory of Riemann surfaces.
Nevertheless, the critical exponents and the leading singular behavior of
the free energy on conical singularities happen to be  closely related  to
that of polyhedra. We will show this in the paper.

Hyperbolic geometry also introduces  an especially interesting type of singularity
which is not sensible for a flat or elliptic metrics. These are hyperbolic  funnels, or  {\it\ cusps}, also called  
parabolic singularities, where singular points are infinitely far away
and do not belong to the surface (Fig.\ref{Fig3}). Formally, cusps can be
seen as a limit \(\alpha\to1\), or \(\g\to 0\) of a conical singularity on
a hyperbolic surface. However, this limit is singular and can not be blindly
taken from formulas for conical singularities.

 Cusps are topologically equivalent to a cylinder. We will see that the finite-size
 energy (in units of central charge)  of the hyperbolic punctured disk (which
comprises half of the pseudosphere in Fig.\ref{Fig3}) with the circumference
\(L\) is equal to that of  a cylinder with the same circumference. If 
\(L=\frac{4\pi}{\log|p|}\) then we will have  
 \begin{align}\la{cusp1}
\text{cusp}:\quad f=-\frac{c\pi}{6L},\quad (p\partial_{p}p) f= - \frac{c}{24}.
\end{align}
We present some results about
critical systems on surfaces with multiple cusps.

The  scaling dimension  with respect to rescaling  the  volume is not the
only critical exponent.  On a surface with multiple  singularities
new exponents emerge. If two  or more singularities merge the free energy scales with the a short distance between the merging singularities and with the large distance if one group of singularities are far separated from another group. 

{ Among many critical exponents we are primarily concerned with two distinguished exponents. One corresponds to a merging of two singularities, another to a separation of one singularity from an aggregate of others.
A genus-0
surface with a discrete set of $n$ curvature singularities  is described
by the Riemann sphere $\hat{\mathbb{C}}$. We denote the coordinates of singularities by  $p_1,p_2,\dots,p_n$. 

 Then, if two singularities  with angles \(\g_{k}\) and \(\g_{j}\)  merge,
the free energy scales  as \begin{align}
 (p_k-p_j)\p_{p_k}f\big|_{p_k\to
p_j}=\Lambda_{kj} .\la{500}
\end{align}
But if 
one conical  singularity, say with
the angle \(\gamma_k\), is separated from others by a large  distance $d(p_k,\{\bf p\})$, this configuration can be also viewed as a result of merging  \(n-1\) singularities, all except one. Then the  free energy scales as \begin{align}
p_k\p_{p_k}f\big|_{d(p_k,\{\bf p\})\to\infty}=-\Delta_k.\la{400}
\end{align}
 Other exponents correspond to merging three or more cones.   
We compute   the two exponents (\ref{500},\ref{400}) and discuss the relations between them, assuming that all conical singularities have generic angles, i.e. the result of merging is a generic conical singularity again. This means that a sum of any number of angles is not an integer. This condition excludes cases when the result of the merging is either a cusp or an orbifold, or both. In this case the critical exponent of multiple merging can be computed in a similar manner.   

Symbolically the  free energy can be represented
as a string of  primary operators \(V_{\al_k}(p_k)\) located
 at \(p_k\) on a surface
without  singularities\begin{align}
e^{-f}\propto\<\prod_mV_{\al_m}(p_m)\>.
\end{align}This interpretation  has been realized by Knizhnik \cite{Knizhnik1987}
for   a flat branched Riemann surface with
identical singularities.

In this representation the dimensions   \(\Lambda_{kj}\) appear as the holomorphic conformal
dimensions  and the operator product expansion (OPE) \begin{align}
\prod_mV_{\al_m}(p_m)\sim |p_k-p_j|^{2\Lambda_{kj}}\prod_{m\neq k,j}V_{\al_m}(p_m) \end{align}
If the result of the merging of two cones is a cusp or an orbifold, the OPE consists of more terms.

 In the case of a punctured sphere, there is only one exponent   \eq{cusp1}
 \begin{align}
(p_k-p_j)\p_{p_k}f\big|_{p_k\to
p_j}=\frac c{24}.\la{600}
\end{align}
This is due to  a  special property of cusps: merging  two or more cusps
is again a cusp. We will extensively explore this property.
In this case,  the OPE possesses descendants which contribute the  logarithmic
corrections to the free energy.}

Perhaps the simplest example of a critical system in 2D is the free boson,
whose free energy is  given by the regularized spectral determinant of the
Laplace operator.  Since, in the critical regime the finite-size part of
the free energy is universal, we can thus identify it with   the logarithm
 of spectral determinant up to terms independent of positions of singularities
  \begin{align}\la{4}
 f=-\frac{c}{2}   \log{\rm Det}'\left(-\Delta\right)+\text{metric independent
terms}.
\end{align}
 This formula leaves the details of the specific critical system  to the
central charge, while dependence of geometry is captured by the spectral
determinant. The determinant is primed to indicate that zero modes have been
excluded from the determinant. For one and two  conical singularities, the
determinant
can
be  computed by various methods  \cite{Dowker_1994,Spreafico_2005,Spreafico_2007,Klevtsov2016-1}.

{  Hence our results for critical exponents could be understood as the limiting
behavior of the spectral determinant as singularities  merge. In a more formal language  we will obtain the limiting behavior of the spectral determinant  on the boundary of the moduli space \(\mathcal{M}_{0,n}\).
}

\subsection{Adiabatic quantum states  and quantum Hall effect}

These ideas and results from critical systems have recently found new life
in the study of adiabatic quantum states. By an adiabatic quantum state,
we mean a system which remains in its instantaneous ground state under the
adiabatic evolution of the parameters of the system. Varying these parameters,
such as the gauge field and metric, will deform the ground state, but will
not drive transitions to excited states. 
As a  specific example, consider  model fractional quantum Hall wave functions.
 These are states   which evolve adiabatically under slowly changing  parameters
of the system. Crucially, the evolution of the states
does not drive transitions to excited states.  

Specifically, the position of the singularities, as well as the degree of
the singularity (e.g. the opening angle of the cone), can serve as the adiabatically
varying parameters. For instance, a singularity can be made sharper, or it
can be braided around another singularity. This is the subject of {\it\ geometric
adiabatic transport}, and will be the main focus of this paper.

 The object of interest in  adiabatic quantum states  is the holonomy, or
the phase the  states acquire when the adiabatic parameters are taken along
a closed contour. Adiabatic transport along non-contractible contours in
parameter space are of a special interest. In this case, the adiabatic phases
are topological in nature and are related to quantized transport coefficients.

  We will consider the geometric transport of quantum Hall states on a  sphere
with elliptic (cone) and parabolic (cusp) singularities. The space of parameters
in this case is naturally identified with the moduli space $\mathcal{M}_{0,n}$
of an $n$-marked  or punctured sphere (see Sec.\ref{moduli-space} for a
definition and
further references), which is isomorphic to the configuration space of the
coordinates of the singularities.


On a sphere the state $| \Psi \rangle$ is not degenerate, and the holonomy
is just a phase. Symbolically, the adiabatic phase reads
\begin{align}
\Phi_\Gamma=\oint_\Gamma\mathcal{A},\quad \mathcal{A}=\ii \<\Psi|\dd \Psi\>,\la{5}
\end{align}
where \(\mathcal{A}\) is the adiabatic connection for the quantum state \(|\Psi\>
\), and the integral goes along a closed path in the space of parameters.

The integer and fractional quantum Hall effects are the most  studied examples
of adiabatic quantum systems, with a body of knowledge that goes nearly as
deep experimentally as theoretically. Quantized {\it electromagnetic} adiabatic
transport has been measured in
quantum Hall (QH) states to metrology precision \cite{Klitzing}. 

Importantly, QH states do not possess conformal symmetry. In contrast to
 critical systems, QH states
feature a scale given by the magnetic length \(\ell=\sqrt{\hbar/eB}\). However,
QH states are holomorphic in the space of complex adiabatic  parameters,
and it is through this property that they are
connected to  conformal field theory (CFT). We will clarify the relation
between QH states and CFT in this paper. 

Holomorphic adiabatic states are states which depend holomorphically on the
complex valued adiabatic parameters (not coordinates of particles). The holomorphic
property is the governing
property
which in the end reduces the problem to conformal field theory.
Specifically, we will show that the operators representing singularities
 transform as conformal primaries.  
The holomorphic  property endows  states with robust transport characteristics.
This observation links geometric transport to critical systems and conformal
field theory. This is the major  result of this paper.
We also conjecture that this is a general property of any {\it holomorphic
adiabatic state}, not just QH states.
 For a recent review on developments in the geometry of QH states
we suggest \cite{Klevtsov:2016_lectures}.

As indicated in \cite{KW2015,Can2016}, the adiabatic phase has an  intensive
part which is {\it topological in nature}, whose contribution depends only
on the homology class of the path \(\Gamma\) in the parameter space, not on its
shape or area. They
are the analog of {\it
theta terms} of quantum field
theory, elusive but important.
This part of the  phase  survives as a non-contractible contour $\Gamma$
is shrunk to zero area about special points in parameter space, and   is
directly
related to precisely quantized
transport coefficients.   It is in the  focus of this paper.  

We will show that (the topological part of) the  adiabatic phases of geometric
transport of QH states  are directly related to the conformal dimensions
of a critical system discussed in the previous Section. 

Specifically let us denote by \({\bf p}=p_1,\dots,p_n\) the complex coordinates
of the singularities and move in such manner that the conformal factor of
the
metric and the magnetic potential are kept fixed (see a detailed definition
in Sec.\ref{Sec4}). Then we
show that the adiabatic connection differs from \(\dd_p f\) by an exact form which does not contribute to the phase
\begin{align}
\mathcal{A}_p=
\dd_p f+\text{exact form.}\la{A}
\end{align}
 Here   $\dd_{p}$ is the Dolbeault operator (see Sec. \ref{connection}
for
a definition), and   \(f\) is the finite-size
free energy  of the  
critical
system which corresponds to the QH state. 

Equivalently this means that the adiabatic phase with respect to a path 
$\Gamma$    in the moduli space is \begin{align}
\Phi_\Gamma=-\IM\oint_{\Gamma} \mathcal{A}_p=-\IM\oint_{\Gamma} \dd_{p}f.\la{7}
\end{align}
 We will also show that the adiabatic phases \eq{7} (in units of  \(2\pi
\))
and appropriate contours \(\Gamma\) are the conformal  dimensions  \eq{400}
and \eq{500}.
Specifically, if \(\Gamma\)  represents a rotation of the entire system about
a singularity with the cone angle \(\g\), then \(\Phi_\g=-2\pi\Delta_\g\),
and if the adiabatic process moves a singularity \(\g\) around a singularity
\(\g'\), then the exchange phase is \(\Phi_{\g\g'}=2\pi \Lambda_{\g\g'}\).

The formula \eq{7} establishes a formal relation between   QH states and
conformal field theory.
It allows to read the  adiabatic phase of geometric transport from the finite-size
energy
of the corresponding  critical system and 
to apply the methods of conformal field theory to interference and transport
 phenomena in a broad  class of quantum systems.  

This relatively new application of conformal field theory  addresses richer
physical phenomena 
than
just thermodynamics of  critical systems.

The formula \eq{7} also defines the notion of `central charge',   a   transport
coefficient \(c\), for  states which generally are not conformal. In the
case of Laughlin's series of states  with the filling fraction
\(\nu\) and spin \(j\) (see Sec. \ref{Sec6.1} for the definition of spin
in QHE),  the   corresponding critical system 
 is characterized by the central charge \begin{align}
c=1-\frac {12}\nu(\12-j\nu)^2.\la{8}
\end{align}

Geometric adiabatic transport of QH states on  smooth surfaces with genus
two or higher has been  discussed in \cite{Levay1997,KW2015}.  The geometric
transport on a torus, the genus one surface, is yet to be studied. In this
paper  we consider the geometric transport on a genus zero surface. In this
case,  the complex structure moduli are constructed from the positions of
singularities. We look  at adiabatic  phases when the  surface is deformed
in such manner that positions of singularities move along  non-contractible
closed paths encircling  other singularities. 
Treating positions of singularities as adiabatic parameters  was suggested
 in recent papers \cite{Can2016,Gromov2016,Can2017}.

 In addition to the quantized transport coefficients, the adiabatic phase
describes the response  to small  changes of geometry. Moving singular points
 around forces the electronic fluid to gyrate. It changes the angular momentum
of the fluid and exerts a torque. These aspects have been discussed in \cite{Can2016,Can2017}.
We do not discuss it here.

The response of QH states to smooth changes of geometry, as well as the effects
of the gravitational anomaly in the QHE have been discussed in many recent
papers \cite{clw,Abanov2014,framinganomaly,Klevtsov2014,CLWBig, Read2015,
lcw,KMMW2016,Klevtsov:2016_lectures}.
{ In contrast to these papers we focus on singularities.}

 Conical singularities naturally appear in some physical settings.  Disclination
defects in a  crystalline lattice are equivalent to conical singularities,
  and they are common in   graphene  \cite{graphene1}. There, two
disclinations with the degree \(\alpha=\pm 1/3 \) are pentagons  and
heptagons, respectively, embedded into the honeycomb lattice of graphene.
{ Conical geometry has also been simulated in photonic systems \cite{schine2015synthetic}.} Cusps can be seen as contact leads, whose end points extend to infinity
and thus are excluded from the sample  \cite{avron_pnueli1992,pnueli1994}.

We mention the early paper
on Landau levels in a singular geometry \cite{Furtado} and recent papers
on the subject \cite{Gromov2016,
Klevtsov2016-1,Klevtsov2016-2, Can2016,Can2017} and the study of QH electric
transport on a surface with a cusp \cite{avron_pnueli1992,pnueli1994}. 
\bigskip

Similar  to the finite-size energy   in critical systems, the   adiabatic
phases do not depend on details of the system. But they also do not depend
on  details of  geometry of the surface. Conformal transformation of the
metric  leaves the adiabatic phase invariant, and we can consider motion in
the space of metrics in a fixed conformal class.  The
 Uniformization Theorem 
asserts that every surface is conformally equivalent to a surface with constant
curvature: positive, negative, or zero. What remains of the space of metrics
is the so-called Teichmuller space of the complex structure moduli \(\mathcal{M}_{0,n}\),
which
is  finite dimensional. In light of these facts, we consider constant curvature
surfaces with a finite number of prescribed singularities, with special attention
paid to negative curvature surfaces. 

The purpose of this paper is  to give an expository   account of the  geometric
transport of QH states on \(\mathcal{M}_{0,n}\), the
moduli space of a sphere with \(n\)-singularities (cones and cusps). This
involves some review and rephrasing of the main results of the previous works
\cite{Can2016,Can2017}, as well as some novel generalizations.

\section{Main results and organization of the paper}\la{2}

The central formula \eq{7} requires us to make a detour into the separate
topic of critical systems on singular surfaces, so that we may determine
the finite-size correction and compute the adiabatic phases. Results in this
domain are scattered and not easily adaptable to our purposes. A part of
this paper is devoted  to discussing conformal field theory on
singular surfaces, and could be read independently from the part on QHE.
 
\bigskip

The purpose of this paper is thus two-fold:\begin{itemize}
\item[ 
 (i)] to connect geometric transport of holomorphic adiabatic states (QH
states in particular) to critical systems  (Sec.
\ref{Sec3}-\ref{Sec5}), and to express topological adiabatic phases in terms
of conformal dimensions of primary CFT\ operators;  
\item[(ii)] to compute dimensions (critical exponents) of singularities in
critical
systems on a singular surface (Sec. \ref{Sec6}, \ref{Poly}).
curvature
transport of QH states.
\end{itemize}

Having established item (i), the details of geometric transport can be extracted
from item (ii), which concerns only critical systems. 

 We deal exclusively with the Laughlin wave function,
though we try to present our results in a manner that makes potential generalization
to other states possible. 

Before turning to the QHE, we start with a general  discussion of holomorphic
adiabatic
states and  introduce the generating
functional in Sec. \ref{Sec3}.  We assert that the generating functional
is quasi-primary  with respect to the M\"obius transformation of the positions
of the singularities, and  that their  dimensions determine
the  topological parts of the adiabatic phase.

In Sec. \ref{Sec4}, we turn to the example of QH states, reviewing the
construction of holomorphic wave functions on a Riemann surface with curvature
singularities. 
Then in Sec. \ref{Sec5}, we use the vertex construction for the Laughlin
wave function and  obtain the central charge \eq{8} of the  critical system
corresponding to
Laughlin's QH states.
In this part, we introduce  the Quillen metric for the fractional QHE generalizing
the Quillen metric for the integer QHE of Refs. \cite{Avron1994, Levay1997,KMMW2016}.

Finally, in Sec. \ref{Sec6}, we   compute the dimensions of conformal field
theory on a 
sphere with singularities. We connect the dimensions to the asymptotes
of accessory parameters and what we call (for lack of an existing name) ``auxiliary"
parameters of the uniformization theory of singular surfaces. We present
the explicit calculation of the accessory and auxiliary parameters and the
free energy of critical system on polyhedra in  \ref{Poly} and    review
the metrics of singular surfaces of revolution in \ref{A1}.  
\ref{Sec8.6}.

\bigskip

The building block of  adiabatic phases is a local  angular momentum of the
quantum state about a given singularity.  Electrons located at the vicinity
of a singularity  gyrate faster if the curvature is positive (or slower if
the curvature is negative) than the bulk electrons. As it was shown in \cite{Can2016},
the excess (or the deficit)
of the angular momentum of a conical singularity (in units
of the Planck constant \(\hbar  \)) is \begin{align}
 {\rm cone}: \quad {\rm L}_{\gamma}= \frac {c}{24}\frac{h_\g}{\g}=\frac {c}{24}\left(\frac 1\g-\g\right). \la{171}
\end{align}
This result is closely related to the formula of Cardy and Peschel  \eq{cardy-cal}.

 If we replace  the  rescaling of the volume $V\p_V$ by the dilatation
operator \(\12 L\p_L \), where $L$ is the linear scale system, and further replace it  by a holomorphic complexification $L \to Le^{i \theta}$
and
identify  $-i \partial_{\theta}$ with the rotation operator. 

The result for the cusp is shown in  \cite{Can2017} to be given
by\begin{align}
  {\rm cusp}:\quad {\rm L_{\rm cusp}}=
\frac{c}{24}.\la{172}
\end{align}
 We will show how to obtain these formulas in Sec. \ref{Sec5.3}.

 We mainly focus on surfaces with a constant curvature $R_0$, but some formulas
remain valid for surfaces with variable curvature. In this case $R_0$ will
be the mean curvature. We denote
 \be \al_0=(V/2\pi)R_0,\la{1111}\ee  
 and call it a `background charge'. It is zero
for polyhedra surfaces. 
 
 If all singularities are conical (a marked sphere), the surface is compact
and
  $\al_0=2(\chi-\sum_j\al_j),$
  where \( \chi\) is the Euler characteristic (\(\chi=2\) for the sphere).
Throughout the paper   we label orders of conical  singularities (the deficit
angles in units of \(2\pi\)) as \(\al_1,\dots,\al_n\), the opening angles
(in units of \(2\pi\)) as  \(\g_1,\dots,\g_n\), the dimensions in \eq{cardy-cal}
as, \(h_1,\dots,h_n\)
, and the angular momenta \eq{171} as \({\rm L}_1,\dots,{\rm L}_n\). 

 The first adiabatic phase \(\Phi_k\) appears when we rotate the entire system
  about a chosen conical singularity, say
\(p_k\), or equivalently rotate a chosen conical singularity around a conglomeration
of the other cones. We can obtain it by studying the scaling behavior (at
a fixed
conformal factor and magnetic potential, see Sec. \ref{Sec6.1}) when one
singularity is sent to
infinity $p_{k} \to \infty$, while the others stay fixed. The result is 
\begin{align}
        {\rm cone}:\quad\Phi_{k} = - 2\pi  \Delta_{k}, \quad \Delta_{k} 
= - \alpha_{k} \sum_{j = 1}^{n} {\rm L}_{j}+\12\al_0 {\rm L}_k.\la{122}
\end{align} 
This
 phase is due to the geometric analogue of the Aharanov-Bohm phase
in which  a particle with  an angular momentum ${\rm L}_{k}$  taken along
a closed path  picks up a phase proportional to the enclosed curvature $\frac{1}{2}{\rm
L}_{k}\times\text{enclosed curvature}$.  The $k^{\rm th}$ singularity will
encircle the total curvature  $(V/4\pi)R_0 - \alpha_{k}=\12\al_0-\al_k$, since it will not
see its own $\alpha_{k}$ curvature. Furthermore, during this process the
other singularities of angular momentum $L_{j}$ will encircle the curvature
$\alpha_{k}$, but in the opposite orientation. These effects together give
\eq{122}.

Another adiabatic phase occurs upon braiding of singularities. Concretely,
this is the process in which two singularities exchange their positions.
 Assume that we can deform the
surface in such manner that  a conical point  with the deficit
angle  \(\alpha_k \) adiabatically encircles  another
conical point with a deficit angle \(\alpha_j\) around an infinitesimal 
small circle. Also assume that \(\al_j+\al_k<1\) (this condition excludes
the case when  the result of merging two conical points is a cusp).  We will
see that  the state acquires the phase 
\begin{align}
\text{cones}:\  \Phi_{jk}=2\pi \Lambda_{jk},\quad \Lambda_{jk}= -\frac{c}{12}\alpha_k\alpha_j+\al_j{\rm
L}_{k}+\al_k{\rm L}_{j}
= \frac{c}{24}\alpha_k\alpha_j\left(\frac 1{\g_j}+\frac 1{\g_k}\right)\la{999}
\end{align}
This result was found in \cite{Can2016}. The last two terms in \eq{999} represent
the geometric analog of the  Aharonov-Bohm (AB) phase, in which a particle
with spin \(L_j\)
(or \(L_{k}\)) encircles a curvature singularity with total integrated curvature
\(4\pi \al_j\;\)(or \(4\pi \al_k\)). 
The first term \( -\frac{c}{12}\alpha_k\alpha_j\)
 (in units of \(2\pi )\) is the braiding phase. It gives the mutual (or exchange)
statistics  to  conical singularities. When they exchanging places, the state
acquires a  phase equal to $-\frac{\pi c}{12}\alpha_k\alpha_j$ plus the AB phase.

The two types of  adiabatic phases are connected  via the sum rule
\begin{align}\la{1001}
\Phi_{k} =      \sum_{j \ne k}^n \Phi_{jk}-2\pi c_1(p_k), 
\end{align} 
where $c_1(p_k)$ is the first `Chern number'  on  the moduli space  equal
to the integral of the adiabatic curvature over the $k^{\rm th}$ hyperplane
of 
the moduli space (excluding its boundary). It is computed in Sec.\ref{chern-class}.
The result  is
\be\la{1609}
\text{cones}:\qquad c_1(p_k)=\frac{c}{24}\al_0\alpha_k.
\ee
The relation between the dimensions then is
the sum rule\begin{align}
\Delta_{k} = - \sum_{j \ne k}^n \Lambda_{jk}+\frac{c}{24}\al_0\alpha_k.\la{1002}
\end{align} 
 This sum rule could be interpreted as follows. The integral of  the  adiabatic
curvature over the compactified moduli space  \(\overline{\mathcal{M}}_{0n}\)
which includes its  boundary  \(\{p_k=\infty,\
 p_k=p_j\}\) vanishes  { (mod \(2\pi\))}. In this integral the contribution
of the boundary
of the moduli space \(\p\mathcal{M}_{0n}\) is \(\Phi_{k}-\sum_{j \ne k}^n
\Phi_{jk}\). It is  balanced by
the  the total  adiabatic phase over
the moduli space \(\mathcal{M}_{0n}\) which  excludes the  boundary. This
is the first Chern number $c_1(p_k)$. One can interpret $c_1(p_k)$ as an
exchange phase between a $k$th particle and the background charge $\al_0$.

{ Now let us turn to a punctured sphere. In this case,  there is only one independent phase,
the braiding  phase \eq{92}, due to a special property of cusps:
merging two or more cusps is again a cusp. Hence when we rotate the system about
\(p_k\) the acquired phase  \(\Phi_k\)  is  \(2\pi\) times the angular
momentum of one cusp \({\rm L_{\rm cusp}}\).
 In \cite{Can2017} it was shown that the angular momentum is also the
braiding  phase of two cusps 
\begin{align}
\Lambda_{\rm cusps}={\rm L_{\rm cusp}}=\frac{c}{24}.\la{92}
\end{align}
 This result does not follow adiabatically as a limit of sharpening
angle of a hyperbolic cone in \eq{999} 
 \(\al\to 1\) (see. Fig.\ref{Fig3}). 
This is an important conclusion. The process  \(\alpha\to 1\), or \(\g\to
0\) on a hyperbolic surface which formally yields to a cusp is not adiabatic.
{  For instance, as a hyperbolic cone is sharpened to become a cusp, the localized fraction of particles at the cone tip are removed suddenly from the system when $\alpha = 1$. }
}


These are some results we obtain in the text of the paper. They have applications
for the limiting  behavior of  the  free energy of a critical system (\ref{500},\ref{400},\ref{600}), due to
 the relation \eq{4}. 

{ The  overall rescaling  \({\bf p}\to
\lambda^{1/2} {\bf p}\)  on a marked  surface with a constant curvature   is equivalent to 
 the rescaling of the  volume  \(V\to\lambda^{-(2-\al_0)}
V\). The rescaling yields the dimension \be -\sum_k p_k\p_{p_k}f
= \Delta_0+\sum_k\Delta_k,\quad \Delta_0=-\frac{ c}{24}\al_0(2-\al_0),\ee 
where $\Delta_0$ is the dimension of the background charge. 

With the help of the identity  $\sum_k\Delta_k=- (2-\al_0)\sum_k{\rm L}_k$ due to \eq{122}, we obtain 
\be  -(V\p_V) f= \frac{ c}{24}\al_0+\sum _k{\rm L}_k.\la{2209}\ee   
This formula generalizes the known results for particular  surfaces with constant curvature: regular surfaces of arbitrary genus \eq{109}, the polyhedra surfaces (see Appendix  \ref{Poly}), the result  \eq{cardy-cal} of \cite{Cardy1988} for  
a single flat
cone, and the results for singular   surfaces surfaces of revolution (see \cite{Klevtsov2016-1} and references therein). 
}

These  results  (due to \eq{4})
 can be expressed  in terms of the limiting behavior of the  spectral determinant
of the Laplace operator near the boundary of moduli space.  At a fixed
conformal factor and regardless from the sign  of the curvature \footnote{ The formula \eq{2509} was known before  for polyhedra surfaces (see Appendix \ref{Poly} and references therein), the formula  similar to \eq{26} for compact surfaces in the  Schottky space was quoted in \cite{zograf1989}.} the limiting behavior of the spectral determinant reads
\begin{align}&
\text{cone:}\quad\log{\rm Det}' (-\Delta)=\begin{cases} p_k\to p_j:\quad\qquad\quad
\frac{1}{6}\alpha_1\alpha_2\left(\frac 1{\g_1}+\frac 1{\g_2}\right)\log|p_k-p_j|, \la{2509}
\\ d(p_k,{\bf p})
\to\infty:\quad\ \frac{1}{6}\left(\al_k
\sum_{j} \frac{ h_{j}}{\gamma_{j}} -\12\al_0 \frac{ h_{k}}{\gamma_{k}}
\right)\log
|p_k |.
\end{cases}\\
&\text{cusp}:\quad 
 \log{\rm Det}' (-\Delta)|_{p_k\to p_j}=\frac{1}{6}\log|p_k-p_j|.\la{2609}
\end{align} 
\bigskip
Also we rewrite the scaling formula   \eq{2209} for the surface with conical singularities as
\be
-(V\p_V) \log{\rm Det}' (-\Delta)=\frac{ 1}{6}\chi+\frac 1{12}\sum_k\left(\sqrt \g_k-\frac 1{\sqrt\g_k}\right)^2.
\ee
\bigskip

  We start from some basic properties of adiabatic holomorphic quantum Hall
states and emphasize their common features with critical systems.
The central property is a transformation law for the adiabatic connection
under $SL(2, \mathbb{C})$ (i.e., a M\"obius transformation of the position
of singularities). This property, plus fusion rules for merging singularities
and the value of the central charge appears to be sufficient to completely
describe the geometric transport.

\section{Geometric Transport of Holomorphic adiabatic states}\la{Sec3}

In this section we introduce the concept of holomorphic adiabatic states
and show that much of the key features of geometric transport follow from
some simple defining properties of these states. Throughout, we have in mind
quantum Hall states as the prototypical example, but we keep the discussion
less specific to stress what we believe is a broader class of many-body quantum
states.  

We begin  with a lightning review of the moduli space of singular
metrics, followed by a discussion of adiabatic transport on such spaces.

This section is the conceptual heart of the paper, with the key concepts
and connections presented, and without derivation. We save the derivation
 to later sections where we deal with 
the   Laughlin's series of QH states. 

\subsection{Moduli space of a sphere with singularities}\la{moduli-space}

We begin with a lightning review of the moduli space of singular metrics,
mainly to introduce nomenclature. For more details, we suggest \cite{hempel1988uniformization,
zograf1988liouville,park2017potentials}. In Sec. \ref{Sec4}, we will describe
the construction of metrics on punctured spheres.

We are primarily concerned with constant curvature $R_{0}$ metrics on genus-0
surfaces with a discrete set of curvature singularities at the points ${\bf
p} = \{p_{1}, ..., p_{n}\}$. Such a Riemann surface $\Sigma$ is described
by the Riemann sphere $\hat{\mathbb{C}}$ with marked or punctured points
${\bf p}$. In the case of  conical singularities, the points belong to the
surface. We refer it as a marked sphere. Cusp points can be viewed as the limit in which hyperbolic cones are sharpened such that their tip is pushed off to infinity. Such a surface
is  non-compact. We refer it as  a punctured  sphere. 

The choice
of complex coordinates is determined by the complex structure moduli, which
define an equivalence class of conformally equivalent metrics. The sphere
has a unique choice of complex structure, so the moduli space is a single
point. The singular sphere, however, has a larger moduli space related to
the space of configurations of punctures $C_{n} = \{ (p_{1}, ..., p_{n}),
p_{i} \ne p_{j}\}$. Upon identifying points which are equivalent under M\"obius
transformations $SL(2, \mathbb{C})$, and permutation of the indices (action
by the symmetric group on $n$ elements ${\rm Symm}(n)$), we obtain the moduli
space $\mathcal{M}_{0,n}$ for the genus-0 n-punctured sphere 
\begin{align}
\mathcal{M}_{0,n} = C_{n} / SL(2, \mathbb{C}) \times {\rm Symm}(n).     
\end{align}
Since a M\"obius transformation can be used to fix the position of three
points on the Riemann sphere, the moduli space ends up having complex dimension
$n-3$. For $n \ge 3$, the uniformization theorem guarantees that there exists
a meromorphic function which maps $\Sigma$ to one of three surfaces $S$:
the sphere ( for $R_{0}>0$), the plane ( for $R_{0} = 0$), or the upper half
plane \(\IM\ w>0\) (for $R_{0}<0$). This map \(w(z): \Sigma \to S\) is known
as the developing map, while the inverse is a covering map often called the
Klein map. The covering map
is always explicitly available for three singularities, essentially due to
the fact that in this case the moduli space  shrinks to a point \cite{kuga1993galois}.
The developing map for $R_{0} = 0$ can also be constructed explicitly \cite{Troyanov2007}
for an arbitrary number of singularities, and is formally identical to the
Schwarz-Christoffel map for polygonal domains.

A non-contractible closed path on a punctured sphere
is not closed in the  \(w\)-plane, but its ends can be brought together by
a modular transformation. These transformations generate the Fuchsian group,
isomorphic to the fundamental group of the punctured Riemann sphere. The
quotient of the upper half
plane with the Fuchsian group is the fundamental domain of the multi-valued
developing map $w(z):\Sigma \to S$.

{  The free energy  at a fixed volume is a regular function on the moduli
space, having
singularities on its boundary. The boundary of the moduli space $\p\mathcal{M}_{0,n}$ corresponds
to configurations where  two or more singular points merge. The moduli space complemented by the boundary  is the compactified  moduli space denoted by $\overline{\mathcal{M}}_{0,n}$. The boundary of the moduli space of a marked sphere consists of points $p_k=p_j,\  j\neq k$ and the $p=\infty$.  In this paper we consider two different boundary configurations: (i) all points merge except one, (ii) 
two points merge.  They correspond to singularities of the free energy expressed
by Eqs.(\ref{500},\ref{400}).
 In the case of a punctured surface the boundary is further reduced. In this case, the boundary of the moduli space is a set of points $p_k=p_j,\  j\neq k$. In this case the  limiting behavior of the free  energy is given by \eq{600}.
}

\subsection{Holomorphic adiabatic states and generating functional}\la{Sec2}

 {The central point of the theory of quantized transport in the    QHE
is that QH states are holomorphic adiabatic states.   

A holomorphic adiabatic state is a holomorphic section of a line bundle fibered
over the space of complex-valued adiabatic parameters. This property  holds
when the adiabatic parameters are magnetic fluxes threading handles of a
multiply-connected surfaces. And  it is also true for geometric transport,
where the adiabatic parameters are complex structure moduli \(\mathcal{M}_{0,n}\).
This seemingly benign definition of holomorphic adiabatic states has profound
consequences, as we will now show.
}

We adopt the inner product of states  with  respect to  a measure \(m(z,\bar
z) dz d\bz\)
 \begin{align}
\<\Psi'|\Psi\>=\int \overline{\Psi'(z_1,\dots,z_N)}\Psi(z_1,\dots,z_N)\,\prod_{i=1}^Nm(z_i,\bar
z_i) dz _id\bz_i,
\end{align} 
chosen such  that it stays unchanged in the adiabatic process. We will specify
the measure for the  states on the lowest Landau level (LLL) in Sec.\ref{Sec6.1}.

Specifically, the  holomorphic state reads
\begin{align}
\Psi(z_1,\dots,z_N|{\bf p})=\frac{\mathcal{X}(z_1,\dots,z_N|{\bf p})}{\sqrt{\mathcal{Z}({\bf
\bar p};{\bf p})}},\la{159}
\end{align} where the non-normalized state \(\mathcal{X}\) is  a multi-valued
holomorphic function  in 
\(\bf p\). The  dependence of anti-holomorphic   \(\bar{\bf p}\) is found
only  in real
normalization factor \(\mathcal{Z}\) 
\begin{align}
\mathcal{Z}({\bf
\bar p};{\bf p})=\int |\mathcal{X}(z_1,\dots,z_N|{\bf p})|^2\,\prod_{i=1}^Nm(z_i,\bar
z_i) dz _id\bz_i.\la{12}
\end{align} 
{ We emphasize that  the  state is holomorphic with respect to the
adiabatic  parameters,  in our case the positions of singularities, and not
necessarily the particle
coordinates.
}
\subsection{ M\"obius transformation of  holomorphic states}\la{SecM}
Adiabatic quantum states on a  closed surface have no physical boundaries,
hence are  invariant under
  a simultaneous M\"obius transformation of coordinates of  particles \(z_1,\dots
 z_N\) and
 coordinates of singularities \(\bf\ p\).  For a finite number of particles, the
numerator and the denominator are  simultaneously M\"obius  invariant. This,
is no longer true when the number of  particles  \(N\) sent to infinity.
In this limit,  the generating functional 
\(\mathcal{Z}\), hence 
the non-normalized state \(\mathcal{X}\), are  transformed under M\"obius
transformation (at a fixed measure \(m(z,\bar z)\)),  in such manner that
the normalized  state \(\Psi\) remains invariant. 

A basic property of such
states which we may take as a definition of holomorphic states is that the
generating functional
transforms as a quasi-primary 

 \begin{align}
p_k\to
g(p_k)=\frac{ap_k+b}{cp_k+d}:\qquad  
\mathcal{Z}(g({\bf p}))=\prod_{k}|g'(p_k)|^{-\Delta_k}\mathcal{Z}({\bf p}).\la{592}
\end{align}
The property \eq{592} appears to be a governing  principle.  We prove it
in Sec.\ \ref{5} for QH states, but would like to emphasize that it represents
a minimal assumption, combined with holomorphicity, which gives rise to conformal
symmetry of adiabatic states. 

 It is tempting to assume that  not just QH states, where we checked  the
transformation  \eq{592}  directly, but a broad class of holomorphic adiabatic
states features a relation to conformal field theory. 

Before discussing the emergent conformal symmetry, we review the consequences
of the  M\"obius transformation on the adiabatic connection, following Ref.\cite{polyakov1970}.

\subsection{Adiabatic Connection }\la{connection}

The holomorphic property has an important consequence. All information of
the adiabatic transport is encoded by the normalization factor \(\mathcal{Z}({\bf
p})\) referred
to
as the {\it generating functional}. Proceeding forward, it is convenient
to express   the adiabatic connection \eq{5} in the holomorphic basis \(\mathcal{A}=
\frac\ii 2 (\mathcal{A}_p+\mathcal{A}_{\bar p})\), where \(\mathcal{A}_p=2
\<\Psi|\dd_p\Psi\>=
\sum_{k=1}^{n}\mathcal{A}_{k} dp_k\) and \(\mathcal{A}_{\bar p}=-2\<\dd_{\bar
p}\Psi|\Psi\>=
\sum_k\bar {\mathcal{A}}_{k} d\bar p_k\), where \be\dd_p=\sum_{k=1}^{n}dp_k\partial_{p_k}\ee
is the Dolbeault operator 
\footnote{Although the sum here extends over $n$ complex dimensions, $SL(2, \mathbb{C})$ symmetry will reduce the number of independent dimensions to $n-3$}.

 A straightforward calculation using \eq{159} shows that the adiabatic connection
is determined by the generating functional \begin{equation}\mathcal{A}_p=
\dd_p\log
\mathcal{Z},\quad \mathcal{A}_{\bar p}=
- \dd_{\bar p}\log \mathcal{Z}.\la{A_connection}\end{equation}
In other words,
the adiabatic curvature is a K\"ahler form, and the generating functional
is the K\"ahler potential. Likewise it follows that the adiabatic phase is
expressed entirely in terms of the generating functional
\begin{align}
\Phi_\Gamma=-\IM\oint_\Gamma \dd_p\log \mathcal{Z.} \la{phase_Z}
\end{align}
For QH states, the K\"ahler property of the adiabatic curvature was known
for a
long time, see e.g.   \cite{Levay}, and has since become standard lore in
the literature. For a more general treatment in the QH setting, along with
a formal proof of this result, see e.g. \cite{KMMW2016}.

 The generating functional contains much more information about the system
than just  the adiabatic phase. Consider for example  conductances associated
with
an adiabatic parameters \(\bf\ p\). According to the Kubo formula  conductances
are
   components of the adiabatic curvature 2-form (see, e.g., \cite{Avron1994})
\be
{ d\mathcal{A}}=\ii \<\bar\dd\Psi| \dd\Psi\>=\sigma_{p\bar
p}(\frac\ii 2 dp\wedge d\bar p) \la{3109} \ee We see
that 
the generating functional describes  the conductance matrix  \be\sigma_{k\bar
l}=\p_{p_k}\p_{\bar p_l}\log\mathcal{ Z}.\la{3209}\ee A   regular part of
the generating functional yields an exact part of the
adiabatic
connection 1-form, which  does  not contribute to the adiabatic phase. This
part,
however, contributes to the conductance, and  as suggested in Ref.\cite{Avron1994}
could  be regarded as mesoscopic
fluctuations. 

Equipped with the adiabatic connection, we note that normalized holomorphic
adiabatic states satisfy
\begin{align}
\left(\partial_{\bar{p}}- \frac{1}{2}   \mathcal{A}_{\bar{p}} \right)\Psi
 = 0.
\end{align}
This property could also serve as a definition of holomorphic  adiabatic
states

\subsection{Quantized transport and  topological part of the adiabatic  phase}\la{Sec3.3}

  The adiabatic
phase    consists of 
 two distinct contributions, a {\it geometric} and
a {\it  topological} part. The geometric part depends on the shape of the
path, whereas the topological part depends  only on the homology of the path.  

The two kinds of phases  can be distinguished by their adiabatic curvature.
The geometric part of the adiabatic curvature is a smooth function of adiabatic
parameters. In contrast,  the topological part 
is accumulated  on  a finite set of  points where the adiabatic curvature
\eq{3109} is a delta-function.  These points occur on the boundary of the
moduli space,
 when  singularities merge. The  weights 
of the delta functions are  quantized conductances \eq{3209},   intrinsic
characteristics
of the quantum state.
 They    can not change continuously and, therefore, are not affected by
small perturbations.  

We focus on the topological part of the phases.
They are determined by the limiting behavior  of the adiabatic connection  \begin{align}\mathcal{A}_k\big
|_{d(p_k,{\bf p})
\to\infty}\sim \frac {\Phi_k}{  2\pi p_k},\quad \mathcal{A}_k\big|_{p_k\to
p_j}\sim  \frac {\Phi_{kj}}{2\pi(p_k-p_j)}.\la{6111}\end{align}
One corresponds to the process when the system rotates by \(2\pi\) about a point \(p_k\). The second corresponds to the braiding of singularities
\(p_k\)
and \(p_j\).
In the rest of the paper we derive the formulas for the  adiabatic phases \(\Phi_k\) and \(\Phi_{kj}\) quoted in Sec.\ref{2}.

\subsection{Transformation Law and M\"obius sum rules for Adiabatic Connection}
If  we assume the transformation property  \eq{592}, then  
the adiabatic connection \(\mathcal{A}_p\)
transforms as 
\begin{align}
\mathcal{A}_p&\to \mathcal{A}_p -\frac{1}{2}\sum_{k=1}^{n} \Delta_k \p_{p_k}
\log g' (p_k)dp_k,\la{2677}
\end{align}
Explicitly, the transformation rule for the components of the connection
 reads\begin{align}
\mathcal{A}_k(g({\bf p}))=(ad-cb)^{-1}(cp_k+d)^2\Big(\mathcal{A}_k({\bf p})+{\Delta_k}\frac
c{cp_k+d}\Big).\la{Amobius}
\end{align}
The invariance under    an overall translation of singularities \(
\sum_k\mathcal{A}_k(g({\bf p}))=0\) yields the two additional  
 sum  rules  
\begin{align}
\sum_k\mathcal{A}_k=0,\quad \sum_k(\Delta_{ k}+2\mathcal{A}_kp_k)=0,\quad
\sum_k(\Delta_{k}p_k+\mathcal{A}_kp_k^2)=0.\la{541}
\end{align}

These formulas can be illustrated by the familiar expression of the three
point correlation  functions of quasi-primary operators. If there are only
 three singularities, the sum rules alone  determines the  adiabatic connections
\begin{align}
\mathcal{A}_1=\frac{\Delta_3-\Delta_1-\Delta_2}{2(p_1-p_2)}+\frac{\Delta_2
-\Delta_3-\Delta_1}{2(p_1-p_3)}.
\la{271}
\end{align}
Other components are obtained by a permutation.

Then the generating functional is determined  up to an overall  constant
reads \cite{polyakov1970}
\begin{align}
\mathcal{Z}(p_1,p_2,p_3)=|p_1\!-p_2|^{\Delta_3-\Delta_1-\Delta_2}
|p_2-p_3|^{\Delta_1-\Delta_2-\Delta_3}|p_3-p_1|^{\Delta_2-\Delta_3-\Delta_1}C(\al_1,\al_2,\al_3).\nn
\end{align}
The constant \(C(\al_1,\al_2,\al_3)\) depends on the orders of singularities.


\subsection{Fusion Rules and Dimensions from M\"obius Sum rules}\la{11}

The symmetry under the  M\"obius transformation is  weaker than 
 conformal symmetry, but in practice it helps one compute the dimensions and the geometric transport.

In the case of  more than three  singularities,
 explicit formulas for
the  connection are only available for the flat  polyhedral surfaces (Appendix \ref{Poly}).
However, the
asymptotes as singularities merge can be obtained using the sum rules \eq{541},
along
with some additional assumption about the result
of the  merging.

 When two singularities  \(p_k\)
and \(p_j\) merge, a new singularity occurs. We
denote the dimension of this singularity \(\Delta_{kj}\), and the new adiabatic
connection  as $\mathcal{A}_{kj}$.
The original components of the adiabatic connection $\mathcal{A}_{k}$ and
$\mathcal{A}_{j}$ diverge upon
merging these singularities. However, the first sum rule requires that the
sum \(\mathcal{A}_k + \mathcal{A}_{j} \),  is regular.  The second sum rule
implies that the original connections diverge as
\begin{align}
\mathcal{A}_{k}\Big|_{p_k \to p_j}  =- \mathcal{A}_{j} \Big|_{p_k \to p_j}=
\frac{\Delta_{kj} - \Delta_{k} - \Delta_{j}}{2(p_k - p_j)} \la{merging}.
\end{align}
From which we learn that the mutual statistics is given
by
\begin{align}
 \Lambda_{kj} = \frac{1}{2}\left(\Delta_{kj}-  \Delta_{k} - \Delta_{j}\right)
. \la{38}
\end{align}
Therefore, the mutual statistics is completely fixed by the dimensions $\Delta_{k}$
and the fusion rule which determines $\Delta_{kj}$. Further relations for conical singularities 
 follow from (\ref{122},\ref{999},\ref{1002}).
For example \(\Delta_{kj}=- \sum_{i \ne k,j} (\Lambda_{ik}+\Lambda_{ij})+c_1(p_k)+c_1(p_j)\).
 The sum rules \eq{541} with a combination of fusion rules appear to be useful
for computing the dimensions. We explore similar sum rules in Sec.(\ref{Sec8.2}).

\subsection{Large \(N\) expansion and conformal field theory}\la{SecFinite}
The transformation formula for the generating functional \eq{592} already
implies a connection to conformal field theory. We will now make this connection
precise, and argue that in fact the geometric transport is captured by the
finite-size correction to the free energy of a critical system. Importantly,
this connection appears in the \ \(1/N\) expansion. For a finite number of
particles, the adiabatic connection is a regular function of the adiabatic
parameters. The singularities of the connection, and hence the quantized
transport, are only strictly seen in the limit of large $N$ number of particles.

 The large $N$  expansion of the generating functional of QH states has been
studied in  Refs.  \cite{clw,CLWBig,Klevtsov2014}. It has the form
\begin{align}
\mathcal{Z}=Z_2 ^{N^2}\cdot Z_1^{N}\cdot N^D\cdot e^{f} \cdot (1+\mathcal{O}(1/N_\Phi)),\la{13}
\end{align} 
where \(Z_2\) and  \(Z_1\) are regular functions on moduli space, and \(D\)
is moduli independent,  but universal  \(\al\) dependent factor related to
the dimensions. We will obtain this expansion in the
next Section, where  we will argue  that    \( \log\mathcal{Z}-f\) is a regular
function on moduli space  \footnote{The sign \(+\) in front of $f$ in \eq{13}
is not a misprint. The generating functional
is the inverse of the partition function of the relevant critical system
(see \eq{502}).}.
Hence,  the adiabatic connection \eq{A_connection} is a differential of 
\(f\), Eqs. (\ref{A},\ref{A_connection}), and the phase, Eq. \eq{phase_Z}
read  \begin{align}
\Phi_\Gamma=-\IM\oint_\Gamma
\dd_pf,\la{222}
\end{align}   
as quoted in the introduction \eq{7}.

This observation \eq{13},  could be traced to papers 
  \cite{Jancovici,Zabrodin2006,Klevtsov2013,clw}. It  formalizes the  relation
between QH states   and
 conformal field theory:\begin{itemize}
\item []
 {\it\ the topological part of the adiabatic phase of holomorphic states
in moduli space is determined by scaling dimensions of  the  corresponding
conformal field theory.
}
\end{itemize}
There are several methods to obtain this result for the QH states. The most
powerful
method where every step is under control   is  based on the Ward identity,
developed in Refs. \cite{Zabrodin2006, clw,Can2016,Can2017} (see also
\cite{CLWBig}). Alternative methods are based on collective field theory
approach
\cite{lcw}, and on the vertex construction \cite{Klevtsov2014}, see also
\cite{KW2015}.
We also mention related  approaches of  \cite{framinganomaly} and \cite{Read2015}.
Among
them the  vertex construction  seems the most economical. We adopt it in
this paper. 

Let us assume the formula \eq{222} for now, and walk one  more step before
turning to the specific example of QH states.

\subsection{Adiabatic connection, quasi-conformal map and  the Schwarzian} 
When we move singularities we change the surface  metric. Under a variation
of the metric the free energy of a critical system changes.  The rate of
change is the stress tensor of the critical system. Hence, the 1-form  \(\dd_{\bf p}f\),
could be expressed
through the (holomorphic component of the) stress tensor of the corresponding
critical system. It is known to be  proportional to the Schwarzian of the
metric. In this Section we use these facts to write the connection in terms
of the Schwarzian and a quasi-conformal map describing the displacement of
singularities. The main tool is the simple equation which  connects  the quasi-conformal
transformation of coordinates to the transformation of the position of singularities
 \(\bf\ p\)  (see, e.g., \cite{zograf1988liouville}).

 We start with the   metric  \(ds^2=e^\phi |dz|^2\) expressed in complex
 coordinates. The conformal factor \(e^\phi\) is a function of the positions
 of the singularities \(\bf p\). When we move singularities the metric changes $ds^2\to ds'^2=ds^2+\dd_{\bf p} (ds^2)$.
The new metric is no longer diagonal. Its general form reads  $ \dd_{\bf p}(ds^2)=
e^\phi \left(\dd_{\bf p} \phi |dz|^2+\mu (d\bz)^2+\bar\mu (dz)^2\right)$, where   the differential \(\mu(d\bz)^2\) is harmonic  Beltrami  differential. It obeys the condition \(\nabla_z\mu=0\), where \(\nabla_z=\p_z+\p_z\phi\). 

The new metric can be  brought    into
the diagonal
form  by  an appropriate choice of  coordinates   \(z'(z,\bz)=z+\xi(z,\bz)\) determined by the Beltrami equation $\p_{\bz}\xi=\mu$ and the condition  
 \begin{align}
\nabla_z\xi=\dd_{\bf p}\phi.\la{233}
\end{align}   
 In terms of   a basis of  displacements of
singularities \(\xi=\sum_k\xi_k\,dp_k\), this relation \eq{233} explicitly reads 
   \be\p_{p_k}\phi-\xi_k\p_z\phi-\p_z\xi_k=0.\ee
It  expresses  the transformation of the metric under a motion of singularities. 

Under a general  change of the   metric, which includes a change of the complex structure and also a  change of the conformal
factor, a change of the free energy is expressed through components of the
stress tensor 
\begin{align}
\dd f=-\frac 1{\pi}\int T\mu\  dzd\bz,\quad \dd f=\frac 1\pi\int
\Theta\dd\phi\  dV.   \la{261}\end{align}Here
\(T\) is the holomorphic
component  of the stress tensor,  \(dV=e^\phi dzd\bar
z\) is the volume element, and \(\Theta\) is the trace of the stress tensor.

 In the conformal field theory with the  central charge less than one (the
case
corresponding to the Laughlin states we consider) the holomorphic component
of the  stress tensor is  proportional to the   Schwarzian
 \begin{align}
&T=\frac {c}{12} \mathcal{S}[\phi],\la{2633}\\ 
&\mathcal{S}[\phi]=-\12(\p_z\phi)^2+\p_z^2\phi.\la{263}
\end{align} 
\la{249}
The  trace of the stress tensor is singular at singularities, but away from
singular points it is proportional to the curvature (the trace anomaly)
  
A classical result of Schwarz asserts that the Schwarzian for surfaces
with constant curvature is the meromorphic function with poles of the second
and the first degree at singularities
\begin{align}\la{2581}
\mathcal{S}[\phi]=\sum_j \left[\frac 12\frac{h_{j}}{(z-p_j)^2}+\frac{C_j}{(z-p_j)}\right].
 \end{align}   
Here, \(h_{j}=\al_j(2-\al_j)\) are the dimensions, Eq.\eq{cardy-cal}, and
\(C_j\) are called
{\it accessory parameters}, further discussed in Sec.\ref{Sec8.2}. We comment
on the derivation of this result in Sec.
\ref{Sec8.1}.

This property allows one to reduce the volume integral in the first equation
of  \eq{261} to a
contour integral   encircling 
 singularities\begin{align}
\dd f=
\frac {c}{12}\sum_j \frac{1}{2\pi \ii}\!\left(\12 h_{j}\oint_{\infty}\frac{\xi
dz}{(z-p_j)^2}+C_j\oint_{\infty}\frac{\xi dz}{z-p_j}\right).
\la{41}\end{align}

 Summing up, the  connection  \( \dd f\) is expressed  through  its  central
charge and the data of uniformization theory. These are two independent problems.
We compute the former in Sec.\ref{Sec5.4}, and the latter in Sec.\ref{Sec6}.
\subsection{Scaling}
The second formula of  \eq{261}
can be used to obtain the overall scaling \eq{2209}.  

A uniform change of
the conformal factor   is equivalent to a change of  the volume  \(\int \frac{\delta  f}{\delta \phi } dV=-V\p_V f\).
 Then  Eq.  \eq{261}
yields \(-V\p_{V}f=\frac 1\pi\int
\Theta dV\). The trace of the stress tensor consists of the regular part (the trace anomaly) proportional to
the scalar
curvature
  \be\Theta^{\rm
reg} = \frac{c}{48} R_{0,}\la{5209}\ee and the singular contributions
 supported only at the set of singular points $z = p_j$.
With the help of the  conservation law \(\p_{\bz}
T+e^{\phi }\p_z \Theta=0\)  we express the trace of  the stress tensor through its holomorphic component  \(\frac 1\pi\int
\Theta dV=-\int\p_{\bar z}\left[\frac{e^{\phi(z)-\phi(z')}}{\bar z-\bar z'}\right]
T(z')dz'd\bz'dz d\bz\).
Then using the asymptote  of the metric  \(\phi|_{z\to p_k}\sim -2\al_k\log|z-p_k|\)
(see  \ref{Sec8.1})   and (\ref{263},\ref{2633}) and \eq{2581} we obtain \be\frac
1\pi\int
\Theta dV=\frac1\pi\int
\Theta^{\rm reg} dV+\sum_k{\rm L}_k=\frac{ c}{24}\al_0+\sum_k{\rm L}_k.\la{5309}\ee
Eq. \eq{2209} follows.

We comment, that the term with the accessory parameters in \eq{41} does not
contribute to the scaling. Rather, it  describes
the polarization of the stress tensor \be C_k=(\pi\gamma_k)^{-1}\oint_{p_k} (z-p_k)\Theta dz.\ee

\subsection{Angular momentum}\la{Sec5.3}
Eq. \eq{261} could be interpreted in another manner.   We can consider $\xi(z)$
as a displacement  of a fluid particle located at \(z\).  A choice  \(z'|_{z\to
p_k}=(z-p_k)e^{i\delta\theta}\) or \(\xi(z)|_{z\to p_k}=\ii(z-p_k)\delta\theta\)
represents a local rotation of a fluid particle about  a conical point
\(p_k\) by angle $\delta\varphi=\delta\theta/\gamma_k$ for a cone, or  by
the angle \(\delta\varphi=\delta\theta\) for a cusp. We see it by expressing
the transformation in terms of the developing map (see (\ref{241},\ref{76})
below) in the fundamental domain \(z'|_{z\to p_k}\sim e^{\ii\varphi}\left
(w(z)-w(p_k) \right)^{1/\g_k}\).  The  adiabatic phase obtained under the
rotation   is the angular momentum times the angle of the rotation
${\rm L}_{k}\delta\varphi$. Computing it with the  help of \eq{41}, we obtain
$${\rm L}_{k}\delta\varphi=\delta f=
\frac {c}{24} h_{k}\oint_{p_k}\frac{\delta\theta }{z-p_k}\frac{dz}{2\pi \ii}=
\frac {c}{24} h_{k}\delta\theta.$$ This yields the Eq.(\ref{171}) \({\rm
L}_{k}=\frac {c}{24} \frac{h_k}{\g_k}\).

The same transformation in case of the cusp  reads \(z'|_{z\to p_k}\sim e^{\ii\theta}
e^{2\pi\ii\zeta}\), where \(\zeta\) is  the coordinate  in the upper half
plane (see \eq{687} below). In this case the angle of the rotation is just
\(\varphi=\theta\). Repeating the calculations we obtain  (\ref{172}) \({\rm
L}_{{\rm cusp}}=\frac {c}{24}\).

Eq.\eq{5309}
gives an interpretation of the trace of the stress tensor \(\Theta\) as a density of angular momentum \cite{Can2016}.

\subsection{Chern number: Integrated Adiabatic Curvature}\la{chern-class}

{
In this Section we compute the first Chern number on the moduli space \footnote{More
accurately, the Chern class and Chern number are defined for the bundles
on a nonsingular manifolds, where it is an integer. The moduli space  is
an orbifold with boundary
points. Nevertheless, we still call this topological characteristic the Chern
number. There will be no integer quantization of this number. Rather, the
`Chern number' of an orbifold is a rational number. A rational quantization
places a constraint on the geometry which supports completely filled LLL
and  fractional QH states.
}  of a surface
with conical singularities,
and obtain a sum rule (quoted above in Eq. \eq{1001})  connecting the dimensions
and  the exchange statistics for conical singularities. 

The first Chern number is the  total flux
of the adiabatic curvature over a closed 2-cycle in the moduli space.
Let us first discuss conical singularities. We choose the $k$th singularity and consider the 2-cycle swept out by the
space of the complex parameter \(p_k\). In the case of conical singularities it is a complex hyperplane $M_{k}$,
which excludes the points $\p M_k=\{p_{j},\ j\neq k,\ \infty\}$ occurring
at the boundary of the moduli space. 

The first Chern number \(c_1(p_k)\) is equal to the adiabatic curvature integrated
over  $M_{k}$.    It picks only the geometric
part of the adiabatic phase. However, since the curvature is a K\"ahler form,
application of Green's theorem relates the first Chern number to the topological
part of the adiabatic phase via    \begin{align}
 c_1(p_k)=\sum_{j\neq k}\Lambda_{kj}+\Delta_k.\la{51c}
\end{align}  Here, we compute the Chern number by integrating the adiabatic
curvature over \(M_k\), i.e., computing the total geometric phase,
\begin{align}\la{5109}
c_1(p_k)&= - \frac{1}{\pi} \int_{M_{p_k}} \partial_{\bar{p}_{k}} \partial_{p_k}
f  d p_{k}d\bar p_{k}  \, = -\frac 1{\pi}\int_{ M_{p_k}}   \partial_{\bar{p}_{k}}\left(
 \frac{1}{\pi} \int \Theta \partial_{p_k} \phi\  dV\right)d p_{k}d\bar p_{k}.
\end{align}
To get the second equality we have utilized the variational formula for the
free energy  Eq. \eq{261}.

The contribution of the singular part of the trace of the stress    does not enter \eq{5109},
since singular points are excluded from the integral over $p_k$.  
 Then keeping only the regular part of the trace of the stress tensor $\Theta^{\rm reg}$, and exchanging
the order of integrals, we are left to evaluate 
\(
\int dV \Theta^{\rm reg} \int_{M_{p_k}}  \partial_{\bar{p}_{k}}
 \partial_{p_k} \phi d p_{k}d\bar p_{k}  \,    . 
\)
The integral over $M_{p_k}$ is dominated by the singularity at $p_{k} \to
z$, where the metric approaches $\partial_{\bar{p}_{k}} \partial_{p_k} \phi
=  -\pi \alpha_{k} \delta(p_k - z) $. Other singularities of $\partial_{\bar{p}_{k}}
\partial_{p_k} \phi$ occur when $p_{k} = p_{j}$, which are excluded from
the domain $M_{p_k}$. Hence
\begin{align}
c_{1}(p_k) =  \frac{\alpha_{k}}{\pi}   \int_{\mathbb{C}/\{\bf p\}}      \Theta^{\rm
reg} dV = \frac{\alpha_{k}}{\pi} \frac{c}{48} R_{0} V = \frac{c}{24}\al_0\al_k.\la{53c}
\end{align}
Combined with \eq{51c}, we reproduce \eq{1001} and \eq{1609} quoted in Sec.\ref{2}.

 We comment that in the case of cusps the $p_k=\infty$ does not belong to the boundary of  $M_k$. In this case \eq{51c} reads $c_1=(n-1)\Lambda_{\rm cusp}=(n-1)\frac{c}{24}$  \cite{Can2017}.

\section{Quantum Hall states on a singular surface }\la{Sec4}
Here we give a brief account of states in the lowest Landau level (LLL)
on  curved surfaces.  For more details we refer to \cite{CLWBig},
and for a mathematically oriented reader we recommend  \cite{Klevtsov:2016_lectures}.

\subsection{Lowest Landau levels on curved surfaces \la{Sec6.1}}

  Electrons reside on a surface  threaded by a uniform  magnetic field
\(B\) normal to the surface  (\(eB>0) \). In units \(e=\hbar=1\) the total
flux through the surface  is   \(N_{\Phi} =  B\, V /(2\pi)\).
 We will define   the  magnetic potential $Q$ as the solution
of the Poisson equation \(2B=-\Delta Q\). 

We write the state   in complex coordinates, where the metric is diagonal
\begin{align}
ds^2=e^\phi|dz|^2,\nn
\end{align}
and use  the transverse gauge, where  the gauge potential  reads
\begin{align}
A_z=\12\ii\p_zQ, \quad A_{\bz}=-\12\ii\p_{\bz}Q.\nn
\end{align}
The magnetic field is the Laplacian of the magnetic potential 
\begin{align}B =- 2 e^{-\phi}\p_z\p_{\bz} Q.\la{507}\end{align}
The spin \(j\)-states in the  LLL with \(N\) particles  are tensors of rank
 \(j\)
 defined
as zero modes of the
anti-holomorphic  momentum operators  on the space of \(\mathcal{J}\)-differentials
\begin{align}
\overline{\nabla}^{(j)}_{i}\Psi_{j}(z_1,\dots, z_N)=0,\qquad i=1,\dots, N.\la{509}
\end{align}
The d-bar  operator in complex coordinates  and with respect to the quantum
mechanical measure $L^2$, reads
  \begin{align}\overline{\nabla}^{(j)}=e^{-\frac j2\phi}(\p_{\bz}-\ii
A_{\bz})e^{\frac {j}2\phi}.\nn\end{align}  
The spin of the state could be an integer
or half integer. For QH states the spin  has been introduced in \cite{CLWBig}
and is an important characteristic of states, often omitted in the literature.

A general solution of the set  of equations \eq{509} on a genus zero surface
reads
\begin{align}\la{567}
 \Psi(z_1,\dots,z_N)=\mathcal{Z}^{-1/2} \mathcal{X}(z_1,\dots,z_N) \prod_{i=1}^N\exp
\left(\12
Q(z_i,\bz_i)-\12j\phi(z_i,\bz_i)\right),
\end{align}
where \(\mathcal{X}\) is a symmetric or antisymmetric holomorphic polynomial,
$\mathcal{Z}$ is the normalization factor.

 The  normalization condition involves  the volume integral over the Riemann
surface \(dV=e^\phi dzd\bz\), and results in the normalization factor \eq{12} \begin{align}
\mathcal{Z}=\int |\mathcal{X}(z_1,\dots,z_N)|^2 \prod_{i=1}^Nm(z_i,\bar z_i)dz_id\bar
z_i,\la{5709}
\end{align}
where the measure 
\begin{align}\la{50}
m=\exp \left(
Q-(j-1)\phi\right),
\end{align}
defines the inner product of holomorphic sections $\mathcal{X}$.

The polynomial \(\mathcal{X}\) has further constraints. Convergence of the
integral \eq{5709} limits the number of admissible particles.  Let us denote
by \(h_N\) the degree of  the polynomial. The polynomial grows as  \(\mathcal{X}\sim
z_1^{h_N}\) as a given variable, 
\(z_1\), tends to infinity.  The growth must be compensated by the measure
  \eq{50}. The conformal
factor and the magnetic potential behave as \(\phi\sim  -4\log |z|,\;\;Q\sim
-2N_\Phi\log|z|\), so the measure \eq{50} falls as \(z_1^{-2(N_\Phi-2
j)}\).        Therefore, the state
can be   normalized if \(h_N-N_\Phi+2j\leq
0\). { This is the Riemann-Roch-Hirzebruch condition which limits the
number of holomorphic sections of the Riemann surface. }

A physical assumption is that the
degree \(h_N\)      grows with \(N\) at most linearly \(h_N=\beta N+{h_0}\),
where \(\beta\) and  \(h_0\) are integer valued parameters. This follows
from the assumption that the interaction between particles is pairwise. Therefore,
the largest admissible
number of particles is the integer part of  \(\beta^{-1}(-N_\Phi + 2j+h_0)\).
This is the scenario we consider. The offset \(h_0\) is called the shift \cite{wen1992shift}.

The maximal admissible state uniformly occupies the surface with a density
\( N/V\approx \nu (N_\Phi/ V)=\nu(eB/2\pi\hbar)\). The parameter \(\nu\equiv
\beta^{-1}=N_\Phi/N\) is the filling  fraction. Such a state  has  no boundaries,
and   therefore  is  invariant under the M\"obius transformation of the Riemann
sphere.  

The M\"obius invariance imposes strong restrictions on admissible polynomials
\(\mathcal{X}\). 
One series of states, the Laughlin states, is
singled out by the condition  \(h_N=\beta(N-1)\). In this case  the
number of admissible  particles is\begin{align}
\beta( N-1)=N_\Phi-2j,\la{211}
\end{align} provided that \(N_\Phi-2j\) is a multiple of \(\beta\). This
condition
is fulfilled by  the polynomial \begin{align}
\mathcal{X}(z_1,\dots,z_N)=\prod_{i>j}(z_i-z_j)^\beta.\la{21}
\end{align}
At \(\beta=1\), the completely filled LLL is the 
Slater determinant of single-particle states.

Under M\"obius transformation \(z_i\to g(z_i),\;p_k\to g(p_k)\) the polynomial
\eq{21} and the measure \eq{50} transforms as 
$$\mathcal{X}\to\prod_i[g'(z_i)]^{\12\beta (N-1)}\ \mathcal{X},\quad \prod_{i}
mdz_i d\bar{z}_i\to\prod_i[g'(z_i)]^{
-N_\Phi+2j} \ \prod_{i} mdz_i d\bar{z}_i.
$$
They compensate each other under the condition \eq{211}.

In order to further specify the form of the wave function, we need more knowledge
of the magnetic potential $Q$ and the metric $\phi$ on singular  surfaces.
We discuss it in the next Section.
\subsection{Metric of a  sphere with singularities}\la{Sec509}

In Sec.(\ref{moduli-space}), we discussed the moduli space of singular metrics.
Here we review the construction of such metrics using the tools from uniformization
theory. 

We refer a  Riemann sphere $\hat{\mathbb{C}} = \mathbb{C} \cup \{\infty\}$
with a set of marked points ${\bf p}=\{p_{1}, ..., p_{n}\}$ with a concentration
of curvature (conical singularities) as a marked sphere \(\Sigma\), and a
punctured sphere  $\Sigma= \hat{\mathbb{C}} /{\bf p}$, where the points \(\bf
p\) are removed. The singularities of a punctured sphere are cusps.

 In complex  coordinates,
where the metric is diagonal \(ds^2=e^{\phi}|dz|^2\),   the one-component
Ricci tensor \(R_{z\bz}=\frac 14e^{\phi}R=-\p_z\p_{\bz}\phi\) for the marked
 surface
with singularities of order \(\al_1,\dots,\al_n\)  reads
\begin{align}
4R_{z\bz}=e^{\phi}R_0+4\pi\sum_k\alpha_k\delta^{(2)}(z-p_k). 
\la{741}
\end{align}
The curvature of the  punctured sphere is  given by the same formula with
\(\al_k=1\). Away from the singularities, the scalar curvature \(R_0\) is
a smooth function. 

Surfaces
with given singularities are conformally equivalent to that of constant curvature.
As we mentioned  above,   adiabatic phases depend only on the conformal classes,
so  it is sufficient to   consider only surfaces with a constant curvature:
 positive, negative and zero.  We normalize the curvature to be \(R_0=0,\,\pm
2\). 

 Metrics of surfaces with constant curvature are  solutions of the Liouville
equation with
prescribed behavior
at  singularities\begin{align}
4\p_z\p_{\bz}\phi=-{R_0}e^\phi,\quad \phi |_{z\to p_j}\sim
-\al_k\log |z-p_j|^2.\la{95}
\end{align}
  We assume that infinity is a regular point without any singularities.

The solution of the Liouville equation
can be formally written in terms of a developing map \(w(z):\Sigma \to S\)
from $\Sigma$ to the fundamental domain 

\begin{align}e^{\phi
} =\frac{4 |w'(z)|^{2}}{(1 +\frac 12{R_0}|w(z)|^2)^{2}},\la{27}\end{align}
with a prescribed behavior
at singular points  followed  from  \eq{95}. For $R_{0} = +2$, the fundamental
domain is a finite region of the Riemann sphere. For $R_{0} = 0$, it will
be a polygon on the complex plane. For $R_{0} = -2$, the range of the developing
map is a finite volume region on the Poincare disk.
 
Hyperbolic surfaces with \(R_0=-2\)  are also represented by the Poincare
metric on
the upper half plane. In coordinates $\zeta=-\ii\frac{ w+\ii}{w-\ii}$, the
metric \eq{27} reads
\begin{align}ds^2=
\frac{|d\zeta|^2}
{\left(\IM\zeta\right)^2},\quad \IM\zeta>0. 
\la{687}\end{align}  
For parabolic (cusp) singularities, it turns out that the metric \eq{687}
is more convenient. 

If 
 \(\alpha <1\), the singularity  is  conical. Locally it is equivalent to
 an embedded
cone
with the
apex angle  $2\arcsin\gamma$, where \(2\pi\gamma=2\pi(1-\alpha)\)
is called a cone angle.   Non-convex surfaces  can contain cone points with
order
$\alpha <0$.  The branch point of a multi-sheeted Riemann surface can be
described by a local metric with negative integer $\alpha$.

An especially interesting case occurs when $\gamma$  or
$1/\gamma$ is an integer. In this case, the puncture is  an orbifold point,
a fixed point  of the action of a discrete group of automorphisms
\cite{Thurston1998}. Though interesting and worth noting, this fact does
not ultimately make a difference in our final results.  

The
Gauss-Bonnet formula for a compact surface of a
constant curvature implies that the integer valued Euler characteristic
 receives local contributions  $\alpha_{k}$ from each  singularity
\begin{equation}
\chi= \frac{V}{4\pi}   R_{0} + \sum_{k} \alpha_{k},  \la{chiS}
\end{equation} The Gauss-Bonnet formula   limits the  total order of   singular
points \(\sum_k\al_k\)  if the volume is finite. If
the  curvature \( R_0\) is positive,  then  \( \sum_{i} \alpha_{k}<
\chi\).  If  all conical points are  sharp \(\alpha>0\)
they exist only on  a sphere, where \(\chi=2\).       
In this case, the degrees of the singularities are restricted by the condition
\( 0<2 -
\sum_{k} \alpha_{k}<2\, {\rm min} (1,\,{\rm min}\,\g)\), where ${\rm min}\g$
is the smallest cone angle $\g_k$ \cite{Troyanov1989,Luo_1992}.
 For example, unless \(\alpha\) is a negative  integer, the only  spherical
 surface with two  isolated singularities  is the spindle, where conical
points are
necessarily the same degree  and   antipodal 
\cite{Troyanov1989} (see Appendix \ref{A1}).

In the case of the   surface with negative   curvature  \({ R}_0
< 0\),  the bound  is reversed \(\sum_{k} \alpha_{k}>\chi \) (we
refer to \cite{hempel1988uniformization,TZ2002} for a review of hyperbolic
geometry). In this case
the
number of cones  or cusps is  limited only from below. On a punctured sphere,
where  all singularities
are cusps $R_0V=4\pi (2-n)$. This condition excludes a
pseudosphere,  a hyperbolic surface of revolution with two singularities
and an edge (see \ref{A1}).

A flat (\(R_0=0\)) compact surface is  a polyhedron, and can be constructed by gluing together flat
triangles. The
vertices of the polyhedron are conical singularities with a conical angle equal to the  sum
of
the
angles of the triangles adjacent to the vertex.    The total degrees of  singularities is
equal
to Euler characteristic \(\sum_k\alpha_k=\chi\). 

Polyhedra of genus zero are the only surfaces where the     developing
map   is  explicit beyond three singularities. It takes the form of the Schwarz-Christoffel
conformal map 
\cite{Troyanov2007,Thurston1998}, expressed conveniently in terms of its
first derivative $w'(z) = \partial_z w$ as
\begin{align}
w'(z)=e^{\frac{\phi_0} 2}  \prod_{j=1}^n(z-p_j)^{-\alpha_j},\quad \sum_j\alpha_j=2.\la{78}
\end{align}     The  conformal 
factor $e^{\phi_0}$   in \eq{78} fixes the volume. At a fixed volume the
conformal factor depends on the moduli ${\bf p}$. 
\subsection{QH states on singular surfaces of constant curvature} The  Laughlin
state \eq{50} for genus-zero surfaces with a constant 
curvature are explicit in terms of the developing map \(w(z)\). In a uniform
magnetic field, the magnetic
potential \eq{507}  can be written explicitly as
 \begin{align}\nn
&Q=-\frac B{2}|w(z)|^2,\quad R_0=0,\quad \text{polyhedra} \\
&Q=- \sign(R_{0}) k\log\left(1+(R_{0}/2)|w(z)|^2\right),\quad R_0\neq 0,\nn
\end{align} where we denoted for \(R_{0}\neq 0\)
\begin{align}
{\rm k}=\frac{4B}{ |R_0|} =\frac{4N_\Phi}{|\al_0|}.\nn
\end{align} 
 Using these formulas we write the most explicit form of the  non-normalized
state  \eq{159}  as a function on the fundamental domain
\begin{align}
\Psi=\mathcal{Z}^{-1/2}\prod_{i<j}\left(z(w_i)-z(w_j)\right)^{\beta}\times
\begin{cases}
e^{-\frac B{2}|w_i|^2}\prod_{k=1}^n(z(w_i)-p_k)^{(j-1)\al_k },\quad
\!\!\!R_0=0,\quad\text{polyhedra}\nn\\
\left(1+|w_i|^2\right)^{-\frac
{\rm k}2+j}z'(w_i)^{j-1},\qquad\quad\ \  R_0=2,\quad \text{elliptic}\nn\\
\left(1-|w_i|^2\right)^{+\frac
{\rm k}2+j}z'(w_i)^{j-1},\qquad\quad\ \  R_0=-2. \quad \text{hyperbolic}\nn
\end{cases}
\end{align}
Despite a lack of
 explicit formulas for  the metric beyond the thrice punctured sphere and
flat polyhedra,
the asymptotes near  singularities  turn out to be sufficient to determine
the
adiabatic phases.

\section{Quantum Hall states and conformal field theory}\la{Sec5}
In this section, we obtain the relations discussed above for QH states. In particular, we prove the connection between geometric transport and critical systems highlighted in \ref{SecFinite}.

\subsection{Vertex construction}\la{Sec4.2}
The vertex construction represents the QH state as an expectation value of
a string of vertex operators of a Gaussian free field. Initially proposed
in \cite{moore1991}, the method was significantly extended  by Ferrari and Klevtsov \cite{Klevtsov2014}, see also \cite{KW2015}.
\({}\)

The electrons in the Laughlin states are represented by the vertex operator
\(e^{\ii X}\) of  the   Gaussian field \(X\) of spin \(j\) with the
charge equal to the filling fraction. The field is  compactified as \(X\sim
X+2\pi\).  Its action reads
\begin{align}
S[X]=\frac \nu\pi \int\left(\frac 1{2}|\p_z X|^2 -\ii j(\p_z\phi\p_{\bz}
+\p_{\bz}\phi\p_{z})
X\right)dzd\bz+\frac{\ii}{2\pi}\int \left(\nu B+\frac 14 R\right)X dV.\la{22}
\end{align} 
The first term in \eq{22}  implies
 the operator product
expansion \(e^{\ii X(z_1)}e^{\ii X(z_2)}\sim |z_1-z_2|^{2\beta}\)  as
\(z_1\to z_2\).
 The second term reflects the spin of the field.  The   last terms describe the coupling to the magnetic field and  curvature. Equivalently, the action may be written
\begin{align}
S[X]=\frac {\nu}{2\pi
}\int |\partial_{z} X|^2 dz d\bar{z}+\frac {\ii}{2\pi
}\int\left (\nu B\!+\12\mu_H R\right)XdV\la{23},
\end{align}where \(\mu_H\), the geometric transport coefficient obtained
by the combination of the second and the fourth terms    
 \begin{align}\quad
\mu_H=\frac 1{2}-{j\nu}\la{26}.
\end{align}
 
We will now show that the correlator of a string of vertex operators \(e^{\ii X}\) localized
at positions of particles \begin{align}
\<e^{\ii\sum_{i=1}^N X(z_i)}\rangle &=\mathcal{Z}_G^{-1}
\int e^{-S[X]}e^{\ii\sum_{i=1}^N
X(z_i)}\mathcal{D}X,\nn
\end{align}
represents  the square of  the amplitude of the non-normalized QH state  (\ref{567},\ref{21}) \begin{align}
\prod_{i=1}^N \<e^{\ii X(z_i)}\rangle dV=|\mathcal{X}(z_1,\dots,z_N)|^2\prod_im(z_i,\bz_i) dz_id\bz_i. \la{25}
\end{align}
Here
\begin{align}
\mathcal{Z}_G&=\int e^{-S[X]}\mathcal{D}X
\la{43}
\end{align}is the partition function of the Gaussian field coupled to a magnetic field.

First we separate the  constant part (the zero mode) \(X_0\) of the field
\(X(z,\bz)=X_0+\tilde X(z,\bz)\), such that  \(\int \tilde X =0\). The
integration over the zero mode gives the condition between the number of
particles and the magnetic flux
\begin{align}N=\nu N_\Phi+\mu_H\chi.\nn\end{align}
 equivalent to \eq{211}.  Here \(\chi=\frac 1{4\pi}\int R \) is the Euler characteristic, and \(\chi=2\)
for the sphere. 

The integration over the remaining modes   \(\tilde X\) gives  \begin{align}\nn
\<e^{\ii\sum_i X(z_i,\bz_i)}\rangle=e^{-\frac {4\pi}{\nu}\sum_{i> j}
\left[\sum_{j\neq i}G(z_i,z_j)+G^R(z_i,\bar z_i)\right]} e^{\sum_{i}Q(z_i,\bar z_i)+\frac{\mu_H}{\nu}\phi(z_i,\bar z_i)}.
\end{align}Here  \(G(z,z')=-\frac{\nu}{4\pi}\<X(z)X(z')\>_{\rm c}\) is the Green function
of the Laplace operator \(-\Delta G=\delta(z-z')- \frac 1 V\), and \(G^R(z)=-\frac{\nu}{4\pi}\<X^2\>_{\rm c}\)
is the regularization of the Green function at merging points defined by using the geodesic distance between the points $d(z, z')$ as the limit   \(G^R= \lim_{z\to z'} \left[G(z, z') + \frac
1{2\pi}\log d(z,z') \right] \). Up to additive constant terms $G^R=\frac{\phi}{2\pi}+\text{const}$. For more details, see \cite{Klevtsov2014}. 
 Comparing with (\ref{50},\ref{21}) we obtain the vertex representation of
the Laughlin $j$-spin state \eq{25}.

Our goal  is to compute the normalization factor of the state
\eq{12}. From \eq{25} it follows\begin{align}
\mathcal{Z}=\int \prod_{i=1}^N \<e^{\ii X(z_i,\bar z_i)}dV_i\rangle=\mathcal{Z}_G^{-1}
e^{\mathcal{F}},
\la{30}
\end{align}
where \begin{align}
e^{\mathcal{F}}=\int \left[\int e^{\ii X(z,\bar z)}dV\right]^Ne^{-S[X]}\mathcal{D}X.\la{47}
\end{align}
\subsection{Quillen metric}So
far the vertex construction is  merely
rewriting  the states in terms of the  Gaussian field coupled to a magnetic field. As such it does not
bring  any additional information.  
The representation became helpful when the authors of  \cite{Klevtsov2014} observed   that   in the  large \(N\) limit the functional \(\mathcal{F}\)
  in \eq{47} is a local functional of the curvature, and therefore, \(d \mathcal{F}\)  is an exact form.
It does not contribute to the adiabatic phase, as \(\oint\dd \mathcal{F}=0\).
A physical reason for this is that \(e^{\ii\tilde X(\xi)}\) does not contain
zero modes, which are solely responsible for adiabatic  phases. 

In the case of the integer QHE (\(\beta=1\)) \(e^\mathcal{F}\) was identified
with the  spectral determinant of the Laplace operator in magnetic field
\(\Delta_B=
(\overline{\nabla}^{(j)})^\dag\overline{\nabla}^{(j)}\),
\begin{align}
e^\mathcal{F}={\rm Det}(-\Delta'_B),\quad \beta=1,\la{577}
\end{align} where the prime indicates that the LLL states are excluded.  This follows from the fermionic version of the integral \eq{47}.
 At \(\beta=1\) the integral \eq{47}
over Bose fields is equivalent to the integral \begin{align}\nn
e^\mathcal{F}=\int e^{-\int \psi^\dag(-\Delta_B')\psi\,}\mathcal{D}\psi\mathcal{D}\psi^\dag,\quad
\end{align} 
over  the Fermi field which does  not contain the modes in the LLL. If \(\psi^{(n)}_k\)
are the wave functions of the \(n\)-th Landau level, and \(E^{(n)}_k\) are their energies,  then the integration in \eq{577} goes over    \(\psi=\sum_{n>
0}\sum_k^{N_n} c^{(n)}_k\psi^{(n)}_k\), where \(c^{(n)}_k\) are Grassmann variables and 
 the  modes in the LLL \((n=0)\) are excluded  \cite{Klevtsov:2016_lectures}: \(e^\mathcal{F}=\prod_{n>0}\prod_{k\leq N_n} E^{(n)}_k\), where \(N_n\) is the number of states on the \(n\)-th level. We comment that in general only the LLL is degenerate on a surface with a finite volume. 

 It is obvious that the spectral determinant of the Laplace operator where
 LLL states are excluded is a regular  function. For smooth surfaces the  expansion of \(\mathcal{F}\)
in gradients of  curvature had been computed in \cite{Klevtsov2013,CLWBig,KMMW2016}.
 
We conclude that \(\log \mathcal{Z}\) and  \(\log \mathcal{Z}_H= \log \mathcal{Z}-\mathcal{F}\)
yield the same adiabatic phase, and that  
  \(\log \mathcal{Z}_H\) 
is equal and opposite in sign to the free energy of the Gaussian field coupled to the
magnetic
field\begin{align}
\mathcal{ Z}_H=\frac{\mathcal{Z}}{e^\mathcal{F}}=
 \mathcal{ Z}_G^{-1}
\la{502}
\end{align}
 In the case of the integer QHE the ratio
 \begin{align}
\mathcal{Z}_H=\frac{\mathcal{Z}}{{\rm Det}'(-\Delta_B)}\nn
\end{align}
is called  the Quillen metric (see \cite{KMMW2016} and references therein).
 The Quillen metric singles out the {\it anomalous} part of the generating functional
 solely responsible for the quantized transport and the generating functional \(\mathcal{F}\) of mesoscopic fluctuations. 
 The advantage of the Quillen metric is that it is exactly computable,
as it is equal to the inverse of the partition function of the free Gaussian
field \eq{502}.  This fact   is referred to as a local index theorem (see e.g.,
\cite{TZlocalindextheorem}).
The ratio \eq{502} extends the notion of the Quillen metric to the  case
of fractional QH states which essentially  differ from Slater  determinants. But, like in the integer case, the ratio \eq{502}  is also  exactly computable.  The formula \eq{502} can be regarded as  the extension of the
Quillen metric and local index theorem  to the fractional QHE.
 In this form it has been introduced in \cite{KW2015}.

  The exact form  
   \(
\dd_p\mathcal{F}
\)  does  not contribute to the adiabatic phase. However,  it contributes
to the conductance \(\sigma_{p\bar
p}=(\sigma_{p\bar
p})_H+\p_{\bar
p}\p_p\mathcal{F}\), where \((\sigma_{p\bar
p})_H=\p_{\bar
p}\p_p\log\mathcal{ Z}_H\). Following  \cite{Avron1994} we interpret
\(\p_{\bar p}\p_p\mathcal{F}\) as  mesoscopic fluctuations
of the conductance, subject to details of the system,  versus  \((\sigma_{p\bar
p})_H\),  a universal    part of the conductance determined solely by geometric
characteristics. 

\subsection{Laughlin states and Gaussian free field} Now let us turn to the partition function of the Gaussian field \eq{43}. It is equal to the inverse of the Quillen 
metric, the only object we need to examine. 

We compute it  by shifting the field \(X=Q+\varphi\). Then the partition function of the Gaussian field is \eq{43} reduces to 
{
\begin{align}
-\log \mathcal{ Z}_G=\frac{1}{2\pi}\int\left(\nu |\p_z Q|^2+\mu_H(\p_z Q\p_{\bz}\phi+\p_{\bz} Q\p_{z}\phi\right)dzd\bz+f,\la{5555}
\end{align}
}
where 
\begin{align}\la{55}
e^{-f}=\int e^{-\frac {\nu}{8\pi}\int(\nabla \varphi)^2dV -\ii\frac {\mu_H}{4\pi}\int R\varphi dV}\mathcal{D}\varphi 
\end{align}
The first factor in \eq{5555} is the extensive part growing with  the number of magnetic flux quanta. It consists of two different contributions.
The first  is of the order of \(N^2_\Phi\). It does not
depends on the metric, hence does not contribute to the adiabatic phase.
  The second term  yields the extensive contribution to the angular momentum
\(\mu_H N_\Phi\) but does not depend on \(p_k\). In a uniform magnetic field it   is simplified
{
\begin{align}
-\log\mathcal{ Z}_G=\frac{ B}{4\pi}\int\left( \nu Q  +2\mu_H 
\phi \right)dV +f\la{578}
\end{align}
}
The  second factor,  is the partition function of the Gaussian field with the background charge \(\mu_H\). It encapsulates  the  geometric transport.  This part  does not depend on the magnetic field, and can be computed independently from the QHE.

If the surface is smooth, \(f\) can be expressed as \begin{align}
&f=\frac{\mu^2_H}{2\pi\nu}\int|\p_z \phi|^2dzd\bz+\frac 12 \log{\rm Det}'(-\Delta).\la{588}
\end{align}
But on  a singular  surface, the first term in \eq{588} 
diverges as a logarithm of  a distance between  singular points when they merge, and requires a regularization. A proper regularization is a  removal of a small ball of the area \(\varepsilon\)     around each singularity, and to ensure that the volume of this ball is the same at all singularities.  This can be done,
of course, but is not necessary  for a  computation of  the adiabatic phase.    To this end, we  need only
the variation of the free energy, or stress tensor. We compute it  in the next section.

\subsection{`Central charge' for Laughlin states}\la{Sec5.4}
As we have seen,  the problem is reduced to the computation of the Gaussian field on a Riemann surface.  The central charge of the Gaussian field  \(c\) could be read  from \eq{55}. It is determined  by the    stiffness \(\nu\), and the background charge. We obtain  \begin{align}
c=1-12\mu_H^2\nu^{-1}.\la{58}
\end{align}
where $\mu_{H}$ is given in \eq{26}. This result is standard. The easiest way to see it is to assume that the surface is smooth and compute the  stress tensor   from \eq{588}.
 We write only the
holomorphic component of the stress tensor.
The contribution of the  term  \(\frac{\mu^2_H}{8\pi\nu}\int(\nabla \phi)^2dV\)
 to the stress is
 \(
T^{(1)}=-\frac{\mu_H^2}{\nu}\mathcal{S}[\phi]. 
\)
The stress tensor  of the spectral determinant  is   controlled by 
 the  gravitational anomaly
 \(
 T^{(2)}=\frac 1{12}\mathcal{S}[\phi]. 
 \)
  Together \(T=T^{(1)}+ T^{(2)}=\frac{c}{12}\mathcal{S}[\phi]\).

The remaining task is to compute the entries of
\eq{2581} and  to evaluate \eq{41}.  These problems belong to the  uniformization theory. 

\subsection{M\"obius transformation and the emergent conformal symmetry}

The vertex construction we have outlined in this section suggests that the generating functional consists of two parts: (i) the partition function of a free field theory coupled to a magnetic field, and (ii) a regular function of the moduli \(e^\mathcal{F}\) \eq{502}.  The partition function of the Gaussian field is further  decomposed into the extensive part proportional to the magnetic field  and the intensive part
 \(e^{-f}\) \eq{55},  the free energy of the critical field theory. Apart from $f$, all contributions are regular functions of the moduli, and consequently do not contribute
to the geometric transport. Hence, we are left with a critical system, or  CFT, on a singular surface. It   fully determines the geometric transport.

We observe  the emergence of global conformal symmetry,  the symmetry with respect to the   M\"obius transformations of coordinates  of  singularities $\bf p$. It follows from the representation of the partition function as a string of primary fields evaluated at the singularity coordinates, in the spirit of the Refs. \cite{Knizhnik1987,Bershadsky1988,calabrese2004}.  The intensive part of the partition function $f$ posses a local conformal symmetry, but  the extensive
term  \(e^{-\frac{ B}{4\pi}\int\left( \nu Q  +2\mu_H 
\phi \right)} \) in 
\eq{55}
and the  term  
\(e^\mathcal{F}\)
do not.
Nevertheless,  they are invariant under the M\"obius transformations. We conclude that  the generating functional is a quasi-primary in the singularity coordinates, as it was stated in Sec.\ref{SecM}.

 \bigskip

\section{Conformal Field theory on a singular surface}\la{Sec6}

Now, we return to the formula \eq{41} expressing the the variation of the
free energy in terms of the quasi-conformal map and the accessory parameters.
To evaluate it we need detailed   information
 about the sub-leading terms of the metric in the expansion about a
singular point \(z\to
p_j\).

\subsection{Sub-leading asymptotes of the metric}\la{Sec8.1}
 The expansion of the developing map around a conical point reads
 \begin{align}w(z)\big|_{z\to p_j}\sim& \left( \frac{z-p_j}{l_j}\right)^{\g_j}\left(
1+\frac{\g_j}{h_{j}}C_j
( z-p_j) +\dots\right), \la{242}\\
 w'(z)\big|_{z\to p_j}\sim&\frac {\g_j} {l_j}\left( \frac{z-p_j}{l_j}\right)^{-\al_j}\left(1+\frac{C_j}{\al_j}
( z-p_j) +\dots\right).\la{241}
\end{align} 
The  scaling coefficient \(l_j\), and the
coefficients
\(C_j\),  are each functions of all singularity coordinates. The coefficients \(C_j\) are called {\it accessory parameters}.

To deal with cusps we use a Poincare mapping of a punctured sphere to the
upper half plane  \eq{687}, (see, e.g.,  \cite{zograf1988liouville}) 
\begin{align}\la{76}
2\pi\ii\zeta(z)\big |_{z\to p_j}\!\!\!\!\sim  \log\left(\frac{z-p_j}{l_j}\right)+C_j
 (z-p_j) \!+\!\dots,\quad 2\pi\ii\zeta'(z)\big |_{z\to p_j}\!\!\!\sim\! \frac
1{ z-p_j}+C_j+\!\dots
\end{align}
The asymptotes  (\ref{242},\ref{241}) ensure that near a conical singularity,
the metric appears locally as
\begin{align}
e^{\phi} |dx_i|^2= \frac{4 \gamma_{j}^{2}|x_{j}|^{-2\alpha_{j}}}{ \left(1 + \frac{R_{0}}{2}
 \left|x_{j}\right|^{2\gamma_{i}} \right)^{2}} |dx_i|^2    , \quad x_{j}
= \frac{z- p_j}{l_{j}}, \label{cone}
\end{align}
and that the limit $\g \to 0$ reproduces the cusp metric for a hyperbolic
surface $R_{0} = -2$  as the punctured disk 
\begin{align}
e^{\phi} |dx_i|^2=      \frac{|dx_i|^2}{|x_{i}|^{2} \log^{2} |x_{i}|} .\label{cusp}
\end{align}
The sub-leading terms  of (\ref{27},\ref{76},\ref{241})
determine the sub-leading asymptotes of the conformal factor \begin{align}\nn
\text{cones}:\qquad&\phi\big|_{z\to p_j}\sim - 2\alpha_{j} \log |z - p_j|
-2\gamma_{j} \log l_{j} + 2 {\rm Re} \left( \frac{C_{j}}{\alpha_{j}} (z -
p_j)\right)  + \dots,\\
&\p_z\phi\big|_{z\to p_j}\sim  -\frac{\al_k}{z-p_j}+\frac{C_j}{\al_j}+\dots\la{361},
\end{align}
The metric for cusps follows simply by the $\g \to 0$ limit of the hyperbolic
cone metric\begin{align}
\text{cusps}:\qquad&\phi\big|_{z\to p_j}\sim  -2\log| z-p_j|-\log\log\frac{|z-p_j|}{l_j}+2\RE\left(C_j(z-p_j\right))+\dots,\nn\\
\p_z&\phi\big|_{z\to p_j}\sim  -\frac{1}{z-p_j}\left(1+\frac 1{\log|\frac{z-p_j}{l_j}|}\right)+C_j+\dots
\la{364}
\end{align} This is to be contrasted with the degeneration of $w(z)$ as $\g
\to 0$, which does not produce $\zeta(z)$. 

 In order to compute the free energy we will need the derivative of the metric
by the position of singularity at a fixed coordinate. For conical points
it  follows from (\ref{361})
\begin{align}
 (\nabla_{p_k}+\delta_{jk}\p_z)e^\phi\big|_{z\to p_j}= 0,\quad \nabla_{p_k}=\p_{p_k}
+2\g_j\p_{p_k}\log l_j.
 \la{367}
\end{align}
The formula for the cusp metric is different. In this case the effect of
the
difference between translation in \(p_k\) and in \(z\)  vanishes as \(z\to
p_k\)\begin{align}
(\p_{p_k}+\delta_{jk}\p_z)\phi\big|_{z\to p_j}=\frac {2\p_{p_k}\log l_j}
{\log|\frac{z- p_j}{l_j}|}\to 0.\la{369}
\end{align}

 In order to clarify  the meaning  of the scales \(l_j\) consider
 small balls of the area \(\varepsilon_j^2\) surrounding the conical
points as a short distance cut-off. Then the cut-off  in the $z$-plane (the
punctured sphere) for each  ball \(|z-p_j|<\delta_j\) is  determined by the
area of each ball $
        \varepsilon_j^2\propto \int_{|z-p_j|<\delta_j} 
e^\phi dzd\bz
 \approx   4\pi \gamma_{j} \left( \delta_{j}/l_{j}\right)^{2\g_{j}}
$.
Thus the scale factor $l_{j}$  relates the cut-off in $z$-plane (uniformization
space)  to the physical cut-off $\varepsilon_j$
 \begin{align}
\varepsilon_j\sim (\delta_j/l_j)^{\g_j}.\la{951}
\end{align} 
In the case of cusps
the relation is \(\varepsilon_j^2\sim1/\log(l_j/\delta_j)\), as it follows
from the formula  for a cone  as \(\g_j\to 0\).

The physically meaningful cutoff requires  $\varepsilon_j$ to be same for
all cones, equal to some short scale cut-off $\varepsilon$ to every singularity,
whereas $\delta_{j}$, a cut-off in the abstracted complex plane may vary.
Hence, $$l_j\sim \delta_j.$$


\subsection{Auxiliary parameters} 
With the help of    formulas  of the Sec.\ref{Sec8.1}  we can solve the 
equation
 \eq{233} for the displacements \(\xi\). Using \eq{361} we find \begin{align}
\xi |_{z\to p_j}=-dp_j-(z-p_j)\dd_p\log l^2_j+\dots\nn
\end{align}  Now we substitute this into \eq{41}, and  use (\ref{367},\ref{369}).
We obtain
the expression for the adiabatic connection in terms of sub-leading metric
 asymptotes
 \begin{align}
\p_{p_k}f=  \frac {c}{12}(C_k+D_k).\la{97}  
\end{align}
where we denoted\begin{align}
D_k=\frac{1}{2}\p_{p_k}\log H, \quad H=\prod_jl_j^{2h_{j}}.
\la{66}
\end{align}

The parameters \(D_k\) play  equally important role as  the accessory parameters.
For lack of a name in the literature, we will call them {\it auxiliary
parameters}.

When all  singularities are cusps,  the function $H$ is identical to  that  defined in
\cite{park2017potentials,takhtajan2017local} as providing a K\"ahler potential for the local
index theorem.

The factor $H$ can be better understood if we express it  with
the
help of cut-offs of singularities \eq{951}
\(\log \sqrt H=\sum_j \frac{h_{j}}{\g_j}\left(\g_j\log
\delta_j-\log\varepsilon_j\right)\).  As we already said, the regularization
of  singularities in  critical phenomena requires that the volume of   the
 cut-off
of different singularities  is the same \(\varepsilon_j={\rm const}\) for
all singularities. In this
case  cut-offs \(\delta_j\)  of different singularities in the uniformization
plane are
 different.  We comment that,   in contrast,
in 2D quantum gravity the cut-off  is uniform in the uniformization plane
\(\delta_j={\rm const}\).

\subsection{Liouville functional}
An important consequence of this formula is that the accessory parameters are generated by the functional $Z_L$  \begin{align}
C_k=\p_{p_k}\log \mathcal{Z}_L.\la{60}
\end{align}
defined via 
\begin{align}
 f=\frac {c}{12} \log(\sqrt{H}Z_L).
\la{119}
\end{align}
 In
\cite{zograf1988liouville,TZ2002}, where the relation \eq{60} has been proven,  $ -2\pi \log \mathcal{Z}_{L}$ is referred
to as the Liouville action or  functional.

\subsection{Metrics on the moduli space }
Each term of \eq{119} which enters the  adiabatic curvature
\begin{align}d\mathcal{A}=  -\frac {c}{12}\sum_{k,l}\left[\p_{p_k}\p_{\bar
p_j}\log Z_L+\p_{p_k}\p_{\bar p_j}\log \sqrt{H}\right] \ii dp_k\wedge
d\bar p_j\la{1141}\end{align} is a  geometric invariant regarded as a K\"ahler
metric on moduli space  \cite{zograf1988liouville,TZ2002}.
The first term  in brackets in \eq{1141} is the  Weil-Petersson metric. The second 
is
called the Takhtajan-Zograf metric. Correspondingly, \(\log Z_L\) and \(\log
H\) are the respective K\"ahler potentials for these metrics on moduli space.
The product  \(Z_L \sqrt{H}\) is related to  the Quillen metric \eq{502}
 \begin{equation}
Z_H=e^{\frac{ B}{4\pi}\int\left( \nu Q  +2\mu_H 
\phi \right)\,}\ (Z_L \sqrt{H})^{\frac{c}{12}.}\nn
\end{equation}

\subsection{Accessory parameters }  \label{Sec8.2}

The Schwarzian \eq{263} of the metric which enters \eq{41}
 is the Schwarz  derivative of the developing map
\begin{align}\mathcal{S}[w]=-\12(\p_z\log w')^2+\p_z^2\log
w'.\la{96} \end{align}
We already know that it is  a meromorphic  function with poles of the second
degree at singularities
(regardless of the  sign of $R_0$). The asymptotes
of the developing  map (\ref{242},\ref{76}), or the  metric (\ref{361},\ref{364})
determine the residues
\begin{align}\la{258}
\mathcal{S}[w,z]=\sum_j \left[\frac 12\frac{h_{j}}{(z-p_j)^2}+\frac{C_j}{(z-p_j)}\right],
 \end{align}  
where the accessory parameters \(C_j\) encode the isometry group of the metric, or equivalently the monodromy group of the developing map. 

It follows from \eq{258} that under   the M\"obius transformation \(z\to
g(z)=\frac{az+b}{cz+d}\)  the
Schwarzian derivative  transforms as  \(\mathcal{S}[w,z]\to
 (g'(z))^2\mathcal{S}[w(g(z)),g(z)]\). 
 Treated as a function of \(z,\ {\bf p}\) and \(C_k\), the Schwarzian \eq{258}
  obeys the property\begin{align}\nn
\mathcal{S}[z;{\bf p}; C({\bf p})]=(g'(z))^2\mathcal{S}[g(z);g({\bf p});
C(g({\bf p}))].
\end{align}
A direct consequence of this is that the accessory parameters transform under
M\"obius transformation as
%
\begin{align}
        C_k(g({\bf p}))=\frac{1}{g'(p_k)}\Big(C_k({\bf p})- \frac{1}{2}{h_{k}}\partial_{p_k}
\log g'(p_k)\Big).\la{552}
\end{align}
This implies that the  1-form $C = \sum_{k} C_{k} dp_{k}$ transforms as
\begin{align}
C\to C - \frac{1}{2} \sum_{k} h_{k} d_{p_k} \log g' (p_k)    
\end{align}
in a manner similar to the adiabatic connection \eq{2677}. 

Consequences of the M\"obius symmetry are described in  Sec.\ref{SecM}. 
We relist them here. 
\begin{enumerate}
\item [(i)]
 The Liouville  functional 
\({Z}_L({\bf p})\)   transforms as a quasi-primary field
\begin{align}
 {Z}_L(g({\bf p}))=\prod_{k}|g'(p_k)|^{-h_{k}}{Z}_L({\bf
p}), \la{59}
\end{align} 
\item[(ii)]
 The sum rules
 \begin{align}
\sum_kC_k=0,\quad \sum_k(h_{ \g_k}+2C_kp_k)=0,\quad \sum_k(h_{k}p_k+C_kp_k^2)=0;\la{54}
\end{align}   
 \item[(iii)]The limiting behavior 
   \begin{align}C_j\big|_{d(p_k,{\bf p})
\to\infty}\sim-\frac {\delta_{jk}}{p_k} h_{k},\quad
C_k\big|_{p_k\to p_j}\sim-\frac 12\frac {\lambda_{kj}}{p_k-p_j},
 \quad \lambda_{kj}=h_{k}+h_{j}-h_{kj},\la{394}
\end{align}  
where    \(h_{jk}\) is the dimension of the singularity obtained as a result
 of merging
two singularities\begin{align}
{Z}_L(p_1,p_2,p_3,\dots p_n)\big|_{p_1\to p_2}\sim|p_1-p_2|^{-\lambda_{12}}{Z}_L(p_1,p_3,\dots
p_n).\la{10909}
\end{align} 
\end{enumerate}
Merging  two singularities determines $\lambda_{kj}$.
In the case of merging cones of the order \(\alpha_k\), and \(\alpha_j\),
we obtain a cone of order \(\al_j+\al_k\) assuming that \(\al_j+\al_k<1\).
Then \(h_{kj}=(\al_k+\al_j)(2-\al_k-\al_j)\), and \(\lambda_{kj}=2\al_k\al_j\).

The fusion rule of cusps is different.  Whereas merging two cones $\alpha_{1}$
and $\alpha_{2}$ will produce a new cone  of the order $\alpha_{1} + \alpha_{2}$,
merging two cusps merely yields another cusp. Therefore, in \eq{394} we set
$h_k=1$ and, also, $h_{kj}=1$. Hence, the asymptote of the  accessory parameters
at merging cusps
is half of what follows   from the formula \eq{106} for cones  by setting
 \(\al=1\). Summing up
\begin{align}
\text{cones:}\quad C_k\sim\begin{cases}
      - \frac {\al_k\al_j}{p_k-p_j}, 
      & \text{if}\ p_k\to p_j 
      \\
     -\frac   {h_{k}}{p_k}, 
     & \text{if}\  d(p_k,{\bf p})\to\infty
 \end{cases},\quad \text{cusps:}\quad C_k|_{p_k\to
p_j}\sim
      - \frac {1}{2(p_k-p_j)}. \la{106}
\end{align}
 A more refined asymptote for cusps \cite{zograf1989,hempel1988uniformization}
is
 \begin{align}C_k\big|_{p_k\to p_j}\sim-\12\frac
{1}{p_k-p_j}\left(1-\frac{\pi^2}{\left(\log|p_k-p_j|\right)^2}+\dots\right).\la{log}\end{align}
We
comment that we were able to obtain the limiting behavior of accessory parameters  solely based 
on M\"obius  invariance and fusion properties  of singularities. These arguments   are applicable to a surface  with positive, zero
or negative 
 curvature. For cones the formulas could  be checked against the  explicit
formula
 for  polyhedral surfaces of Appendix \ref{Poly}.

Another comment is that the relation  (\ref{97}) suggests that  the auxiliary parameters 
\(D_k\)  transform under  M\"obius similar to the accessory parameters \eq{552},
and that not only \(Z_L\), but also the  functional \(H\) is a quasi-primary.

The limiting behavior of   accessory
parameters   (\ref{106}) is rather obvious. Nevertheless
it is not easy  to find them in the literature (see however  \cite{zograf1989,hempel1988uniformization,ZZ}). In Sec.\ref{Sec8.6} we give an alternative derivation of Eq. (\ref{106}).

\subsection{Limiting behavior of accessory and auxiliary parameters }\la{Sec8.6}
The accessory and auxiliary parameters  \(C_k\) and \(D_k\) are singular
at the boundary of the moduli space. This means that as singularities merge,
we expect simple poles in \(p_k\)  to appear.  Their asymptotes as singularities
merge
 determine the critical exponents and consequently the topological part of
the adiabatic phase. 
 Here we present heuristic arguments for how to obtain these asymptotes.

We start from cones and assume that the sum of their orders is not a positive
integer.

Consider  merging singularities \((p_k,\al_k)\)
and \((p_j,\al_j\)).  When the points are close, the asymptotes of the metrics
given by \eq{361} \(\phi\big|_{z\to
p_j}\) and \(\phi\big|_{z\to
p_k}\) are valid in a common  domain.  Then the expansions of both  metrics
about
 the middle point \(z=\12(p_j+p_k)+x\)  must give the same result. 
 Hence,
 \begin{align}&\left(\phi\big|_{z\to
p_k}\right)\big|_{z\to
p_j}\sim-2\al_k\log\left|p_j-p_k\right|-2\g_k\log l_k+2\RE x\left(\frac{C_k}{2\al_k}+\frac{\al_j}{p_k-p_j}\right)\nn
\end{align} 
must stay the same under a  permutation of \(j\) and \( k\).
This condition yields 
the asymptotes 
\begin{align}\la{105}
& l_k^{\g_k}
 \sim|p_j-p_k|^{\al_j},\\
\text{cones}:\quad  p_k\to
p_j\qquad &C_k\sim-\frac{\al_k\al_j}{p_k-p_j},\\
&D_k\sim\!\left(\al_j\frac {h_k}{\g_k}+\al_k\frac
{h_j}{\g_j}\right)\!\frac{1}{2(p_k-p_j)}.\la{39}
\end{align}
The limiting behavior of the accessory parameters matches that obtained from
the sum rules in Sec.\ref{Sec8.2} (Eq.\eq{106}). Essentially these calculations
mean that a fusion of two cones of degrees \(\al_k\)
and \(\al_j\) results into the cone of the degree \(\al_k+\al_j\).

The asymptote of \(D_k\) as \(d(p_k,{\bf p})
\to\infty\) follows from the asymptote of
\(C_k\) \eq{106} and the sum rule
\eq{1002}.
 However, it is instructive to obtain it by matching asymptotes.


{  We evaluate the metric at three points. One is $|z|\gg |{\bf p}|$, a regular
 point far separated  from all singularities. The asymptote of the  metric
there  is
  \(\phi|_{z\to\infty}\sim -2(\sum_k\al_k )\log
|z|+\mathcal{O}(z^{-1})\). The
second point is close to   \(z\to p_k\).  The metric  there is  \(\phi|_{z\to
p_k}\sim -2\al_k\log|z-p_k|-2\log
l_k^{\g_k}\). The third point is close to the antipode, where singularities
aggregate. The metric there is  \(\phi_{z\to 0}\sim
-2(\sum_{j\neq k}\al_j)\log|z|-2\sum_{j\neq k}\log
l_j^{\g_j}\). These  asymptotes must match in the common domain
$$\left(\phi|_{z\to\infty}\right)|_{z\to p_k}=\left(\phi|_{z\to p_k}\right)|_{z\to
0}=
\left(\phi|_{z\to 0}\right)|_{z\to p_k}.$$ 
This condition gives  the asymptote of the scaling factors and   the asymptote
of the accessory and auxiliary parameters
 \begin{align}
  & l_{j}^{\g _{j}} \sim \delta_{jk}|p_k|^{\sum_{l\neq j}\al_l}
+(1-\delta_{jk})|p_k|^{\alpha_k}, \\
\text{cones}:\quad d(p_k,{\bf p})\to\infty\qquad &C_j\sim-\delta_{jk}\frac   {h_{k}}{p_k},\la{103}\\
& D_j\sim\frac{\delta_{jk}}{2p_k} \sum_{j\neq k}\left(\alpha_{k}
\frac{h_{j}}{\gamma_{j}} + \frac{h_{k}}{\gamma_{k}} \al_j\right).\la{1039}
\end{align}
The limiting behavior of the parameters  for conical points is well illustrated
 by the explicit formulas for polyhedra surfaces collected in Appendix \ref{Poly}.

Eqs. (\ref{39},\ref{1039}) determines the limiting  behavior, and  transformation properties  of the potential $H$ under the M\"obius transformation (cf.{\ref{59},\ref{10909}) for the Liouville functional)
\begin{align}
& H(g({\bf p}))=\prod_{k}|g'(p_k)|^{\sum_{j\neq k}\al_k\al_j\left(2+
\frac{1}{\gamma_{j}} + \frac{1}{\gamma_{k}} \right)}H({\bf
p}), \la{5909}\\
&H(p_1,p_2,p_3,\dots p_n)\big|_{p_1\to p_2}\sim|p_1-p_2|^{\al_1\al_2\left(2+
\frac{1}{\gamma_{1}} + \frac{1}{\gamma_{2}} \right)}H(p_1,p_3,\dots p_n).\la{10909}
\end{align} 

The short distance asymptote of   auxiliary parameters for cusps can be deduced
from that for cones, by replacing $l_k^{\g_k}\to l_k$ and setting $\al_k=1$.
Then Eq. (\ref{105}) yields
\begin{align}
  & l_{k} \sim   |p_j-p_k|,\\
  \text{cusps}:\quad p_{k}
\to p_{j}\qquad &C_k\sim - \frac {1}{2(p_k-p_j)},\la{118}\\ &D_{k} \sim \frac{1}{p_{k}
- p_{j}}.
 \end{align}
 

These formulas  and the formulas \eq{106} complete the calculations of the
 limiting behavior of the parameters of uniformization theory which  determine
 the dimensions of critical systems  and geometric transport coefficients
of QH states on genus zero singular surfaces.

Combining these formulas with (\ref{97},\ref{6111})
 we obtain the asymptotes of the  free energy  (\ref{500},\ref{400},\ref{600}) and  the adiabatic
phase   (\ref{A},\ref{122})
quoted in Sec. \ref{Sec2},   the dimension and the  OPE exponents (\ref{171},\ref{999})
for cones and cusps  (\ref{172},\ref{92}).
 We see that the the accessory parameters determine the exchange statistics
\(-\frac{c}{12}\alpha_k\alpha_j\), the first term in \eq{999}, and auxiliary
parameters determine the AB phases
\(\al_j{\rm
L}_{k}+\al_k{\rm L}_{j}\) of \eq{999}. In the case of cusps
the exchange phase is half of what would be obtained from cones at \(\al=1\),
due to  the factor \(1/2\) in the asymptote of the accessory parameter
in \eq{106}.

 \section*{Acknowledgements}
We thank  S. Klevtsov and L. Takhtajan for discussions and interest in this
work. The research of
PW was carried out under   support of   IITP RAS  of the Russian Foundation
for Sciences (project 14-50-00150). 
The work of TC was supported   by the Simons Center for Geometry and Physics and 
 in part by the NSF under the  Grants
NSF DMR-1206648.

 \appendix

\section{Singular Surfaces of Revolution}\la{A1}


Here we summarize  the metric of surfaces  of constant curvature with two
punctures.  These are surfaces of revolution immersed in 3D Euclidean space.
In this case the  developing map is either a power law
or a logarithmic function. 

 If the cone is  flat (\(R_0=0\) in \eq{741}), 
a singular conformal
map \(z\to w(z)=z^\g\) brings the metric \eq{27} to the Euclidean form  \begin{align}ds^2=|dw|^2,\quad
w(z) = z^{\gamma}\la{133}.\end{align}
   It maps a punctured disk centered at the conical point to a wedge   of
a plane  \(0\leq{\rm arg}\, w<2\pi\gamma\), whose sides are
isometrically  glued together as  `wedge-periodic' condition (see the Fig.
\ref{Fig1}).
\begin{figure}[htbp]
\begin{center}
\includegraphics[width=15.5cm]{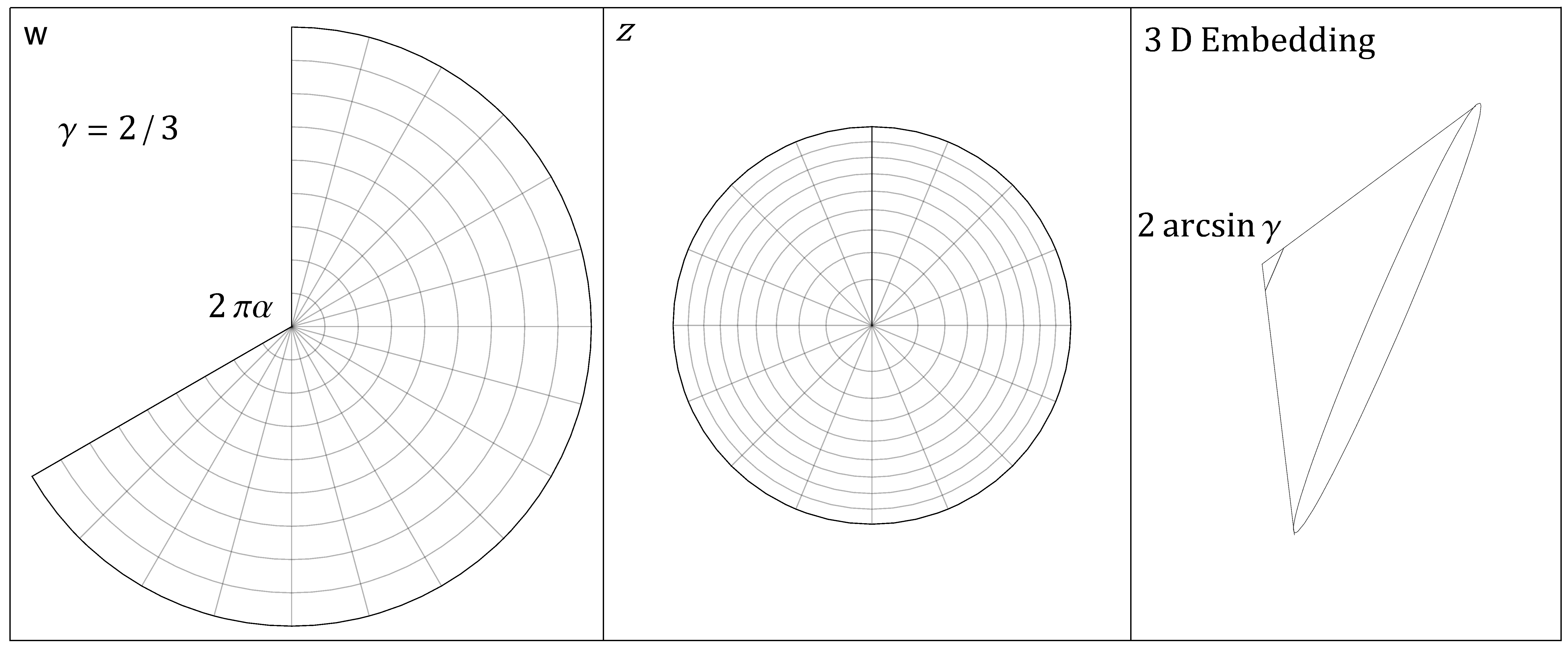}
\caption{Schematic diagram of a cone and its 3D embedding}
\label{Fig1}
\end{center}
\end{figure}
   
  In polar coordinates \(w= \frac{1}{\g}re^{ \ii\gamma\theta}\), where \(\theta\)
runs the full circle (\(0\leq\theta<2\pi \)), and \(r=\g|z|^\g \) the metric
reads\begin{align}
ds^2=r^2 d\theta^2+\gamma^{-2} dr^2. \la{131}
\end{align}

A  model for the  metric with a constant positive curvature \(R_0=+2\)  
is the
spindle, or   
`football',  a surface of revolution of a spherical arch. 
Its metric is the singular conformal map of a sphere with the removed sector
to a twice  punctured plane\begin{align}\la{17}
        ds^{2} =\left(\frac{2|dw|}{1 +|w|^2}\right)^2,\quad w(z)=z^\g.
\end{align}
The spindle has two identical antipodal elliptic cones at \(z=0\) and  \(z=\infty\).
 It is  the only   surface  of constant curvature  with two  isolated singularities
 \cite{Troyanov1989}. 

In polar coordinates \(w=\frac {r}{\ii\sqrt{\g^2-r^2}}e^{\ii\g\theta}\) the
spindle metric reads\begin{align}
ds^2=r^2d\theta^2+\frac{dr^2}{\g^2-r^2}.\nn
\end{align}
Another useful form of the metric 
\(ds^2=dt^2+\g^2\sin^2 t\,d\theta^2\) found in coordinates \(r=\g\sin
t\).  

A model for the negative constant curvature  \(R_0=-2\) are  Minding
surfaces, Fig.\ref{Fig3}.   They are the  surfaces of revolution of  a curve
 \(y(x)\) defined  by the
equation \begin{align}
1+y'^2=\frac{1}{\gamma^2+x^2}.\la{733}
\end{align}
 In polar coordinates the metric of Minding surfaces 
reads\begin{align}
ds^2=r^2d\theta^2+\frac{dr^2}{\g^2+r^2}.  \nn
\end{align}
If   \(\gamma^2<0 \)  the singularity is smoothed.
That curve is called Minding tubular.

If \(\gamma^2>0\) the 
surface has two conical singularities.   \begin{figure}[htbp]
\begin{center}
\includegraphics[width=15cm]{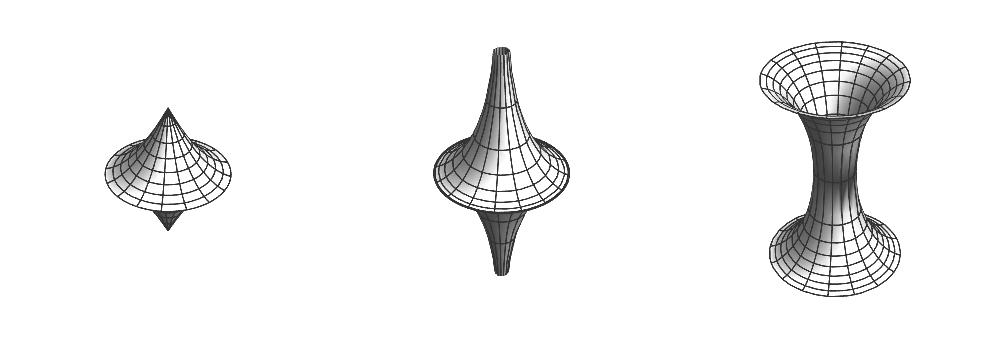}
\caption{Minding surfaces: conical, pseudosphere (middle), and tubular,  
surfaces of revolution with constant negative curvature.}
\label{Fig3}
\end{center}
\end{figure}
These  are the only surfaces  of  constant negative curvature with conical
singularities
  immersed in 3D. 

 Another
form of the metric  of the Minding cones is  \(ds^2=dt^2+\g^2\sinh^2
t\,d\theta^2\) in  coordinates \(r=\g\sinh t\). 

In complex coordinates \(w=\frac {r}{\ii\sqrt{\g^2+r^2}}e^{\ii\g\theta} \)
the Minding cones (\(\g^2>0\)) is a singular map of   the Poincare disk which
removes
a  
sector of the disk \begin{align}
ds^{2} =\left(\frac{2|dw|}{1 - |w|^2}\right)^2,\quad w(z)=z^\g.\la{19}
\end{align}
The map \(\zeta=-\ii\frac{ w+\ii}{w-\ii}\) brings this metric to Poincare
half plane model\begin{align}
ds^2=\frac{|d\zeta|^2}{(\IM\zeta)^2}, \quad \IM\zeta>0.\la{A66}
\end{align}
Conical singularities on surfaces  of negative curvature have an important
degeneration, the cusp, or the parabolic singularity. At \(\alpha\to 1\),
or \(\gamma\to 0\)   
 a hyperbolic    cone degenerates
into a cusp, whose tip is sent to infinity. In this case  the curve \(y(x)
\) \eq{733} becomes the  tractrix 
$$y'=-\frac
1x\sqrt{1-x^{2}}.$$
 Such surface is called   the pseudosphere. As it follows
from \eq{19}, the metric of the cusp  in  polar coordinates   reads \begin{align}
ds^2=r^2 d\theta^2+r^{-2} dr^2\la{20}.
\end{align}
The metric \eq{20} is conformal in the coordinate 
   \( z=e^{-\frac 1 r-\ii\theta}\)
   \begin{align}
ds^2=\frac{|dz|^2}{|z|^2\log^{2}|z|},\quad |z|<1.
\la{141}
\end{align}
 In the coordinate \(\zeta=\ii\log z\) the metric becomes Poincare model
\eq{A66}.

A distinguished feature of  surfaces with constant negative  curvature  immersed
in 3D Euclidean space is an unavoidable edge (a consequence of Hilbert's
theorem:
 Complete isometric immersions
 of surfaces with constant negative  curvature do not exist).  \bigskip

\section{Explicit formulas for polyhedra surfaces}\la{Poly}
 

The explicit formulas for polyhedra surfaces \( \sum_k\al_k=2,\ \al_0=0\)
illustrate the limiting behavior of  the accessory and  auxiliary parameters.

The developing map for a polyhedra surface is  the Schwarz-Cristoffel 
map
\begin{equation}
w'(z)=e^{\frac{\phi_0} 2}  \prod_{j=1}^n(z-p_j)^{-\alpha_j},\end{equation}
where \(\bf\ p\) are images of polyhedra vertices, and  the conformal factor
$\phi_0$ is fixed by the volume
$$  e^{-\phi_0}=\frac 1V\int \prod_{j=1}^n|z-p_j|^{-2\alpha_j}|dz|^2.
$$
The metric is  \(ds^2=e^\phi|dz|^2\) with\begin{align}
\phi=\log|w'|^2=\phi_0-2\sum_j\al_j\log|z-p_j|.
\end{align} Then the explicit formulas for the  accessory and  auxiliary
parameters follow from their definitions (\ref{241},\ref{66})
\begin{align}
&C_k= \partial_{p_k} \log Z_{L} = -\sum_{k\neq j}\frac{\al_j\al_k}{p_k-p_j},
\la{32}\\
&l_j^{-2\g_j}=e^{\phi_0}\prod_{l\neq j}|p_j-p_l|^{-2\al_l}, \la{611}\\
  &D_k=\12\partial_{p_k} \log H = \sum_{j\neq k}
\frac{1}{2(p_k-p_j)}\left(\al_j\frac {h_k}{\g_k}+\al_k\frac
{h_j}{\g_j}\right)-\12 \p_{p_k}\phi_0\sum_j\frac {h_{j}}{\g_j}.\la{61111}
\end{align}
 The Liouville functional $Z_L$, and the functional $H$, the K\"ahler
potentials for the Weil-Petersson and Takhtajan-Zograf metrics on the moduli
space, follow from (\ref{60},\ref{66})
\begin{align}&Z_L\propto\prod_{j>k}|p_j-p_k|^{-2\al_k\al_j}, \la{102}\\
& H=\prod_j l_j^{2h_j}= e^{-\phi_0 \sum_j\frac {h_j}{\g_j}}
\prod_{j>k}|p_j-p_k|^{2(\al_j\frac {h_k}{\g_k}+\al_k\frac
{h_j}{\g_j})}.\nn
\end{align}
The symbol \(\propto\) means `up to moduli independent terms'. These formulas
determine the free energy and the spectral determinant  via the relation \(\partial_{p_i}f = \frac{c}{12} (C_{i} + D_{i}) \).  We write  the free energy  in various suggestive forms
 \begin{align}\la{B7}
e^{-f}\propto\left({\rm Det}(-\Delta)\right)^{-\frac c 2}\propto (Z_L\cdot \sqrt
H)^{-\frac c{12}}\propto &=  e^{\phi_0\sum_k\frac {c}{24} \frac {h_{j}}{\g_j}}
\prod_{k<j}|p_j-p_k|^{-\frac
{c}{12}\al_k\al_j(\frac 1{\g_k}+\frac 1{\g_j})} \\
&=\prod_{j}l_j^{-\frac{c}{12}h_{j}}\prod_{k<j}|p_j-p_k|^{\frac{c}{6}\al_k\al_j}\la{B8}\\
&=e^{-\frac{c}{12}\phi_0}\prod_{j}l_j^{-\frac {c}{12}\al_j}.\la{B9}
\end{align}
Formulas  (\ref{B7}-\ref{B9}) is the result of \cite{Aurell1,Aurell2}, 
 and also  rigorously proven results of \cite{kokotov2013polyhedral,kokotov2011spectral}.
 

Eq. \eq{B9} represents the decomposition \eq{119} of the partition function
on the Liouville-Polyakov action and the functional \(H\).
Eq. \eq{B9} presents the multiplicative form of the partition function
as
a product of scales of individual conical points. 

The formula \eq{B9} is readily
for a generalization for higher genus surfaces. The  first factor 
  there represents the overall scaling \(e^{-\frac {\chi}{24}\phi_0}\) of
surfaces with the Euler characteristic  \(\chi\).  For more details see \cite{kokotov2013polyhedral}.
The most general formula is \eq{2209}. 

 The relation between  the adiabatic connection and the free energy \eq{A}
yields
the application to QHE. The adiabatic connection of Laughlin states in the
moduli space of a polyhedra is
\begin{align}
\mathcal{A}_k=\p_{p_k}f =   \sum_{j\neq k}
\frac{\Lambda_{kj}}{p_k-p_j}-\p_{p_k}\phi_0 \sum_{j} {\rm L}_{j}, \la{1255}
\end{align}
 where  $\Lambda_{kj}=\frac{c}{24}\al_k\al_j(\frac 1{\g_k}+\frac 1{\g_j})$
is  the  mutual statistics \eq{999}, and ${\rm L}_{j} = \frac{c}{24} \frac{h_{j}}{\gamma_{j}}$
is the angular momentum \eq{171}.

We see that the adiabatic connection for the polyhedra is fulfilled by its
asymptotes \eq{6111}. In this   case the parameters \(C\) and \(D\) and the
adiabatic connection are rational holomorphic functions of the moduli and
 the  partition function is the square of a holomorphic function.  This is
no longer
true if the surface is not flat. However, as we  had seen  the limiting behavior
at merging 
 conical points does not depend on the background
metric, and could be read from the flat case.  However, the dimensions $\Delta_k$
\eq{122} explicitly depend on \(\al_0\).  That part is missed in polyhedra.
 Polyhedra are flat surfaces. Their first Chern number vanish \(c_1(p_k)=0\).
The adiabatic phase of  polyhedra possesses only the topological part of the
phase. 

The overall scaling is  encoded  by the first factor in \eq{B7}: because the surface is flat the scaling  ${\bf p}\to\lambda^{1/2}{\bf p}$ at a fixed volume yields  the transformation of the
conformal factor  \(e^{\phi_0}\to \lambda^2 e^{\phi_0}\), or, at a fixed conformal factor to 
the  scaling of the  volume \(V\to \lambda^{-2} V\). Under the scaling 
 the partition function transforms  as \(e^{-f}\to \lambda^{2\sum_j {\rm
L}_j} e^{-f}\).

 If the conformal factor $\phi_0$ is fixed, the partition function
is the quasi-primary
 \be  p_k\to g(p_k):  \qquad e^{-f}\to \prod_j|g'(p_j) |^{-2\Delta_{k}}e^{-f},\la{B11}\ee
where $\Delta_k$ is given by \eq{122}.  

At a fixed volume the partition function is invariant under the   M\"obius
transformation $p_k\to g(p_k)$.
This follows from \eq{1255}, the transformation property of the conformal
factor $e^{-\phi_0}\to \prod_k |g'(p_k)|^{-2\al_k}e^{-\phi_0 }$,  and the
identity \(
 \al_k\sum_j{\rm L}_j=\sum_{k\neq j}\Lambda_{jk},\)
rested on the polyhedra condition $\al_0=0$.  This is the version of the
 sum rule \eq{1002} for the polyhedra surface.

\bibliographystyle{unsrt}

\bibliography{conepaper.bib}

\begin{thebibliography}{10}

\bibitem{Can2016}
T.~Can, Y.{\hspace{0.167em}}H. Chiu, M.~Laskin, and P.~Wiegmann.
\newblock Emergent conformal symmetry and geometric transport properties of
  quantum {H}all states on singular surfaces.
\newblock {\em Physical Review Letters}, 117(26), 2016.
\newblock \href{http://arxiv.org/abs/1602.04802}{\tt arXiv:1602.04802}.

\bibitem{Cardy1988}
J.~L. Cardy and I.~Peschel.
\newblock {Finite-size dependence of the free energy in two-dimensional
  critical systems}.
\newblock {\em Nuclear Physics B}, 300:377--392, 1988.

\bibitem{Knizhnik1987}
V.~G. Knizhnik.
\newblock {Analytic fields on {R}iemann surfaces. {II}}.
\newblock {\em Communications in Mathematical Physics}, 112(4):567--590, 1987.

\bibitem{dixon1985}
L.~Dixon, J.~A. Harvey, C.~Vafa, and E.~Witten.
\newblock Strings on orbifolds.
\newblock {\em Nuclear Physics B}, 261:678--686, 1985.

\bibitem{bershadsky1987}
M~Bershadsky and A~Radul.
\newblock g-loop amplitudes in bosonic string theory in terms of branch points.
\newblock {\em Physics Letters B}, 193(2-3):213--218, 1987.

\bibitem{sonoda1987}
H.~Sonoda.
\newblock Functional determinants on punctured {R}iemann surfaces and their
  application to string theory.
\newblock {\em Nuclear Physics B}, 294:157, 1987.

\bibitem{zamolodchikov1987}
Al~B Zamolodchikov.
\newblock Conformal scalar field on the hyperelliptic curve and critical
  ashkin-teller multipoint correlation functions.
\newblock {\em Nuclear Physics B}, 285:481--503, 1987.

\bibitem{calabrese2004}
P.~Calabrese and J.~L. Cardy.
\newblock Entanglement entropy and quantum field theory.
\newblock {\em Journal of Statistical Mechanics: Theory and Experiment},
  2004(06):P06002, 2004.
\newblock \href{http://arxiv.org/abs/hep-th/0405152}{\tt arXiv:hep-th/0405152}.

\bibitem{Aurell1}
E.~Aurell and P.~Salomonson.
\newblock On functional determinants of laplacians in polygons and simplicial
  complexes.
\newblock {\em Communications in Mathematical Physics}, 165(2):233, 1994.

\bibitem{Aurell2}
E.~Aurell and P.~Salomonson.
\newblock Further results on functional determinants of laplacians in
  simplicial complexes.
\newblock {\em arXiv preprint hep-th/9405140}, 1994.
\newblock \href{http://arxiv.org/abs/hep-th/9405140}{\tt arXiv:hep-th/9405140}.

\bibitem{kokotov2011spectral}
A.~Kokotov.
\newblock On the spectral theory of the laplacian on compact polyhedral
  surfaces of arbitrary genus.
\newblock In {\em Computational Approach to {R}iemann Surfaces}, pages
  227--253. Springer, 2011.

\bibitem{kokotov2013polyhedral}
A.~Kokotov.
\newblock Polyhedral surfaces and determinant of laplacian.
\newblock {\em Proceedings of the American Mathematical Society},
  141(2):725--735, 2013.

\bibitem{Dowker_1994}
J.~S. Dowker.
\newblock Effective action in spherical domains.
\newblock {\em Communications in Mathematical Physics}, 162(3):633--647, 1994.
\newblock \href{http://arxiv.org/abs/hep-th/9306154}{\tt arXiv:hep-th/9306154}.

\bibitem{Spreafico_2005}
M.~Spreafico.
\newblock Zeta function and regularized determinant on a disc and on a cone.
\newblock {\em Journal of Geometry and Physics}, 54(3):355--371, 2005.

\bibitem{Spreafico_2007}
M.~Spreafico and S.~Zerbini.
\newblock Spectral analysis and zeta determinant on the deformed spheres.
\newblock {\em Communications in Mathematical Physics}, 273(3):677--704, 2007.
\newblock \href{http://arxiv.org/abs/math-ph/0610046}{\tt
  arXiv:math-ph/0610046}.

\bibitem{Klevtsov2016-1}
S.~Klevtsov.
\newblock Lowest landau level on a cone and zeta determinants.
\newblock {\em Journal of Physics A: Mathematical and Theoretical},
  50(23):234003, 2017.
\newblock \href{http://arxiv.org/abs/1609.08587}{\tt arXiv:1609.08587}.

\bibitem{Klitzing}
K.~von Klitzing.
\newblock Metrology in 2019.
\newblock {\em Nature Physics}, 13(2):198--198, 2017.

\bibitem{Klevtsov:2016_lectures}
S.~Klevtsov.
\newblock Geometry and large {$N$} limits in {L}aughlin states.
\newblock {\em Travaux Mathematiques}, 24:63--127, 2016.
\newblock \href{http://arxiv.org/abs/1608.02928}{\tt arXiv:1608.02928}.

\bibitem{KW2015}
S.~Klevtsov and P.~Wiegmann.
\newblock {Geometric Adiabatic Transport in Quantum {H}all States}.
\newblock {\em Physical Review Letters}, 115:086801, 2015.
\newblock \href{http://arxiv.org/abs/1504.07198}{\tt arXiv:1504.07198v2}.

\bibitem{Levay1997}
P.~L\'evay.
\newblock {{B}erry's phase, chaos, and the deformations of {R}iemann surfaces}.
\newblock {\em Physical Review E}, 56:6173--6176, 1997.

\bibitem{Gromov2016}
A.~Gromov.
\newblock Geometric defects in quantum {H}all states.
\newblock {\em Physical Review B}, 94:085116, 2016.
\newblock \href{http://arxiv.org/abs/1604.03988}{\tt arXiv:1604.03988}.

\bibitem{Can2017}
T.~Can.
\newblock Central charge from adiabatic transport of cusp singularities in the
  quantum {H}all effect.
\newblock {\em Journal of Physics A: Mathematical and Theoretical},
  50(17):174004, 2017.
\newblock \href{https://arxiv.org/abs/1611.05563}{\tt arXiv:1611.05563}.

\bibitem{clw}
T.~Can, M.~Laskin, and P.~Wiegmann.
\newblock {Fractional Quantum {H}all Effect in a Curved Space: Gravitational
  Anomaly and Electromagnetic Response}.
\newblock {\em Physical Review Letters}, 113:046803, 2014.
\newblock \href{http://arxiv.org/abs/1402.1531}{\tt arXiv:1402.1531v2}.

\bibitem{Abanov2014}
A.~G. Abanov and A.~Gromov.
\newblock {Electromagnetic and gravitational responses of two-dimensional
  noninteracting electrons in a background magnetic field}.
\newblock {\em Physical Review B}, 90:014435, 2014.
\newblock \href{http://arxiv.org/abs/1401.3703}{\tt arXiv:1401.3703v1}.

\bibitem{framinganomaly}
A.~Gromov, G.~Y. Cho, Y.~You, A.~G. Abanov, and E.~Fradkin.
\newblock {Framing Anomaly in the Effective Theory of the Fractional Quantum
  {H}all Effect}.
\newblock {\em Physical Review Letters}, 114:016805, 2015.
\newblock \href{http://arxiv.org/abs/1410.6812}{\tt arXiv:1410.6812v3}.

\bibitem{Klevtsov2014}
F.~Ferrari and S.~Klevtsov.
\newblock {FQHE on curved backgrounds, free fields and large {N}}.
\newblock {\em Journal of High Energy Physics}, 2014(12):1--17, 2014.
\newblock \href{http://arxiv.org/abs/1410.6802}{\tt arXiv:1410.6802v3}.

\bibitem{CLWBig}
T.~Can, M.~Laskin, and P.~B. Wiegmann.
\newblock {Geometry of quantum {H}all states: Gravitational anomaly and
  transport coefficients}.
\newblock {\em Annals of Physics}, 362:752--794, 2015.
\newblock \href{http://arxiv.org/abs/1411.3105}{\tt arXiv:1411.3105v3}.

\bibitem{Read2015}
B.~Bradlyn and N.~Read.
\newblock {Topological central charge from {B}erry curvature: Gravitational
  anomalies in trial wave functions for topological phases}.
\newblock {\em Physical Review B}, 91:165306, 2015.
\newblock \href{http://arxiv.org/abs/1502.04126}{\tt arXiv:1502.04126v2}.

\bibitem{lcw}
M.~Laskin, T.~Can, and P.~Wiegmann.
\newblock {Collective field theory for quantum {H}all states}.
\newblock {\em Physical Review B}, 92:235141, 2015.
\newblock \href{http://arxiv.org/abs/1412.8716}{\tt arXiv:1412.8716v2}.

\bibitem{KMMW2016}
S.~Klevtsov, X.~Ma, G.~Marinescu, and P.~Wiegmann.
\newblock Quantum {H}all effect and {Q}uillen metric.
\newblock {\em Communications in Mathematical Physics}, pages 1--37, 2016.
\newblock \href{http://arxiv.org/abs/1510.06720}{\tt arXiv:1510.06720}.

\bibitem{graphene1}
P.~E. Lammert and V.~H. Crespi.
\newblock {Graphene cones: Classification by fictitious flux and electronic
  properties}.
\newblock {\em Physical Review B}, 69:035406, 2004.

\bibitem{schine2015synthetic}
N.~Schine, A.~Ryou, A.~Gromov, A.~Sommer, and J.~Simon.
\newblock Synthetic landau levels for photons.
\newblock {\em Nature}, 534:671--675, 2015.
\newblock \href{http://arxiv.org/abs/1511.07381}{\tt arXiv:1511.07381}.

\bibitem{avron_pnueli1992}
J.~E Avron, M.~Klein, A.~Pnueli, and L.~Sadun.
\newblock {H}all conductance and adiabatic charge transport of leaky tori.
\newblock {\em Physical Review Letters}, 69(1):128, 1992.

\bibitem{pnueli1994}
A.~Pnueli.
\newblock Scattering matrices and conductances of leaky tori.
\newblock {\em Annals of Physics}, 231(1):56--83, 1994.

\bibitem{Furtado}
C.~Furtado, B.~G.C. da~Cunha, F.~Moraes, E.R.Bezerra de~Mello, and V.B.
  Bezzerra.
\newblock {{L}andau levels in the presence of disclinations}.
\newblock {\em Physics Letters A}, 195(1):90 -- 94, 1994.

\bibitem{Klevtsov2016-2}
D.~Coman, S.~Klevtsov, and G.~Marinescu.
\newblock Bergman kernel asymptotics for singular metrics on punctured
  {R}iemann surfaces.
\newblock {\em arXiv:1612.09197}, 2016.
\newblock \href{http://arxiv.org/abs/1612.09197}{\tt arXiv:1612.09197}.

\bibitem{Avron1994}
J.~E. Avron, R.~Seiler, and P.~G. Zograf.
\newblock Adiabatic quantum transport: Quantization and fluctuations.
\newblock {\em Physical Review Letters}, 73:3255--3257, 1994.
\newblock \href{http://arxiv.org/abs/hep-lat/9405017}{\tt
  arXiv:hep-lat/9405017}.

\bibitem{Note1}
The formula (\ref {2509}) was known before for polyhedra surfaces (see Appendix
  \ref {Poly} and references therein), the formula similar to (\ref {26}) for
  compact surfaces in the Schottky space was quoted in \cite {zograf1989}.

\bibitem{hempel1988uniformization}
J.~A. Hempel.
\newblock On the uniformization of the n-punctured sphere.
\newblock {\em Bulletin of the London Mathematical Society}, 20(2):97--115,
  1988.

\bibitem{zograf1988liouville}
P.G. Zograf and L.A. Takhtadzhyan.
\newblock On {L}iouville's equation, accessory parameters, and the geometry of
  {T}eichm{\"u}ller space for {R}iemann surfaces of genus 0.
\newblock {\em Mathematics of the USSR-Sbornik}, 60(1):143, 1988.

\bibitem{park2017potentials}
J.~Park, L.~A. Takhtajan, and L.-P. Teo.
\newblock Potentials and chern forms for weil--petersson and takhtajan--zograf
  metrics on moduli spaces.
\newblock {\em Advances in Mathematics}, 305:856--894, 2017.
\newblock \href{http://arxiv.org/abs/1508.02102}{\tt arXiv:1508.02102}.

\bibitem{kuga1993galois}
M.~Kuga.
\newblock {\em Galois' Dream: Group Theory and Differential Equations: Group
  Theory and Differential Equations}.
\newblock Springer Science and Business Media, 1993.

\bibitem{Troyanov2007}
M.~Troyanov.
\newblock {On the Moduli Space of Singular {E}uclidean Surfaces}.
\newblock {\em IRMA Lectures in Math. and Theor. Phys.}, 11:507--540, 2007.
\newblock \href{http://arxiv.org/abs/math/0702666}{\tt arXiv:1504.07198}.

\bibitem{polyakov1970}
A.M. Polyakov.
\newblock Conformal invariance of critical fluctuations.
\newblock {\em ZhETF Pis'ma}, 12:538--540, 1970.

\bibitem{Note2}
Although the sum here extends over $n$ complex dimensions, $SL(2, \protect
  \mathbb {C})$ symmetry will reduce the number of independent dimensions to
  $n-3$.

\bibitem{Levay}
P.~L{\'e}vay.
\newblock {{B}erry phases for {L}andau {H}amiltonians on deformed tori}.
\newblock {\em Journal of Mathematical Physics}, 36(6):2792--2802, 1995.

\bibitem{Note3}
The sign \(+\) in front of $f$ in (\ref {13}) is not a misprint. The generating
  functional is the inverse of the partition function of the relevant critical
  system (see (\ref {502})).

\bibitem{Jancovici}
B.~Jancovici, G.~Manificat, and C.~Pisani.
\newblock Coulomb systems seen as critical systems: finite-size effects in two
  dimensions.
\newblock {\em Journal of Statistical Physics}, 76(1-2):307--329, 1994.

\bibitem{Zabrodin2006}
A.~{Zabrodin} and P.~{Wiegmann}.
\newblock {Large-N expansion for the 2D Dyson gas}.
\newblock {\em Journal of Physics A: Mathematical General}, 39:8933--8963,
  2006.
\newblock \href{http://arxiv.org/abs/hep-th/0601009}{\tt arXiv:0601009v3}.

\bibitem{Klevtsov2013}
S.~Klevtsov.
\newblock {Random normal matrices, {B}ergman kernel and projective embeddings}.
\newblock {\em Journal of High Energy Physics}, 2014(1):1--19, 2014.
\newblock \href{http://arxiv.org/abs/1309.7333}{\tt arXiv:1309.7333v2}.

\bibitem{Note4}
More accurately, the Chern class and Chern number are defined for the bundles
  on a nonsingular manifolds, where it is an integer. The moduli space is an
  orbifold with boundary points. Nevertheless, we still call this topological
  characteristic the Chern number. There will be no integer quantization of
  this number. Rather, the `Chern number' of an orbifold is a rational number.
  A rational quantization places a constraint on the geometry which supports
  completely filled LLL and fractional QH states.

\bibitem{wen1992shift}
XG~Wen and A~Zee.
\newblock Shift and spin vector: New topological quantum numbers for the hall
  fluids.
\newblock {\em Physical review letters}, 69(6):953, 1992.

\bibitem{Thurston1998}
W.~P. {Thurston}.
\newblock {Shapes of polyhedra and triangulations of the sphere}.
\newblock {\em Geometry and Topology Monographs}, 1:511--549, 1998.
\newblock \href{http://arxiv.org/abs/math/9801088}{\tt arXiv:math/9801088v2}.

\bibitem{Troyanov1989}
M.~Troyanov.
\newblock {Metrics of constant curvature on a sphere with two conical
  singularities}.
\newblock {\em Lecture Notes in Math}, 1410:296--306, 1989.

\bibitem{Luo_1992}
Feng Luo and Gang Tian.
\newblock {L}iouville equation and spherical convex polytopes.
\newblock {\em Proceedings of the American Mathematical Society},
  116(4):1119--1119, 1992.

\bibitem{TZ2002}
L.~Takhtajan and P.~Zograf.
\newblock {Hyperbolic 2-spheres with conical singularities, accessory
  parameters and {K}{\"a}hler metrics on $\mathcal{M}_{0,n}$}.
\newblock {\em Transactions of the American Mathematical Society},
  355:1857--1867, 2002.
\newblock \href{http://arxiv.org/abs/math/0112170}{\tt arXiv:math/0112170},
  Erratum \url{http://www.math.stonybrook.edu/~leontak/TZ \%20Erratum.pdf}.

\bibitem{moore1991}
G.~Moore and N.~Read.
\newblock Nonabelions in the fractional quantum {H}all effect.
\newblock {\em Nuclear Physics B}, 360(2-3):362--396, 1991.

\bibitem{TZlocalindextheorem}
L.~A. Takhtajan and P.~G. Zograf.
\newblock A local index theorem for families of {$\bar{\partial}$}-operators on
  punctured {R}iemann surfaces and a new {K}{\"a}hler metric on their moduli
  spaces.
\newblock {\em Communications in Mathematical Physics}, 137(2):399--426, 1991.

\bibitem{Bershadsky1988}
M~Bershadsky and A~Radul.
\newblock Conformal field theories with additional {$Z_N$} symmetry.
\newblock {\em International Journal of Modern Physics A}, 2(01):165--178,
  1987.

\bibitem{takhtajan2017local}
L.~A. Takhtajan and P.~Zograf.
\newblock Local index theorem for orbifold riemann surfaces.
\newblock {\em arXiv preprint arXiv:1701.00771}, 2017.
\newblock \href{http://arxiv.org/abs/1701.00771}{\tt arXiv:1701.00771}.

\bibitem{zograf1989}
P.~G. Zograf.
\newblock {L}iouville action on moduli spaces and uniformization of degenerate
  {R}iemann surfaces.
\newblock {\em Algebra i Analiz}, 1(4):136--160, 1989.

\bibitem{ZZ}
Alexander Zamolodchikov and Al~Zamolodchikov.
\newblock Liouville field theory on a pseudosphere.
\newblock {\em arXiv:hep-th/0101152}, 2001.
\newblock \href{http://arxiv.org/abs/hep-th/0101152}{\tt arXiv:0101152}.

\end{thebibliography}

\end{document}